\documentclass[11pt,a4paper]{article}
\pdfoutput=1
\usepackage{amssymb,amsmath,amsfonts, mathtools, mathrsfs, graphicx, caption, subcaption, doi, cite, hyperref, url, dsfont, enumerate}
\usepackage[utf8]{inputenc} 
\usepackage[dvipsnames]{xcolor}

\usepackage{verbatim}
\usepackage{amsfonts,euscript,amssymb,stmaryrd,braket}
\usepackage{tikz}
\usetikzlibrary{arrows,decorations.markings,patterns}
\usepackage{slashed}

\def\clock{{\count0=\time
		\divide\count0 60
		\ifnum\count0<10 0\fi\the\count0
		\multiply\count0 -60 \advance\count0 \time
		:\ifnum\count0<10 0\fi \the\count0
}}
\newcommand{\timestamp}{{\small\vbox{\hbox{\tt\jobname.tex}
			\hbox{\the\day/\the\month/\the\year, \clock}}}}

\makeatletter
\let\old@startsection=\@startsection
\let\oldl@section=\l@section
\renewcommand{\@startsection}[6]{\old@startsection{#1}{#2}{#3}{#4}{#5}{#6\mathversion{bold}}}
\renewcommand{\l@section}[2]{\oldl@section{\mathversion{bold}#1}{#2}}
\makeatother

\numberwithin{equation}{section}

\graphicspath{{figures/}}

\begin{document}
	\renewcommand{\thefootnote}{\arabic{footnote}}

	\overfullrule=0pt
	\parskip=2pt
	\parindent=12pt
	\headheight=0in \headsep=0in \topmargin=0in \oddsidemargin=0in

	\vspace{ -3cm} \thispagestyle{empty} \vspace{-1cm}
	\begin{flushright} 
		\footnotesize
		\textcolor{red}{\phantom{print-report}}
	\end{flushright}

\begin{center}
	\vspace{1.2cm}
	{\Large\bf \mathversion{bold}
	Entanglement entropy and its linear response following a global quench in holographic Gauss-Bonnet gravity}
	\\
	\vspace{.2cm}
	\noindent
	{\Large\bf \mathversion{bold}}

	\vspace{0.4cm} {
		Sabyasachi Maulik$^{\,a,}$\footnote[1]{mauliks@iitk.ac.in} and Soumen Pari$^{\,b,c,}$\footnote[2]{soumen.pari@saha.ac.in}},
	\vskip  0.4cm
	
	\small
	{\em
		$^{a}\,$Department of Physics, Indian Institute of Technology Kanpur, Kalyanpur, Kanpur,\\ Uttar Pradesh 208016, India.
		\vskip 0.1cm
		$^{b}\,$Theory division, Saha Institute of Nuclear Physics,1/AF, Bidhannagar, West Bengal 700064, India.
		\vskip 0.3cm
			$^{c}\,$ Homi Bhabha National Institute, Training School Complex, Anushaktinagar, Mumbai 400094, India.
	}
	\normalsize
	
\end{center}

\vspace{0.1cm}
\begin{abstract} 
	Growth of entanglement entropy in time-dependent states formed due to a global quench in holographic conformal field theories which admit an Einstein-Gauss-Bonnet dual gravity description is studied. The global quench in the bulk is modelled by an AdS-Vaidya solution with an electric charge. It is observed that the Gauss-Bonnet correction parameter leads to faster thermalization, and lower saturation entropy. The rate of growth also depends crucially on the correction, and may exceed $1$ in general spacetime dimensions. Nevertheless, the growth still follows the universal pattern expected for relativistic CFTs. Additionally, a time-dependent analogue of relative entropy introduced in \cite{Lokhande:2017jik} is generalized to include correction from the Gauss-Bonnet parameter. We demonstrate our findings through concrete examples, including instantaneous, linear, and periodically driven quenches. We also briefly mention the evolution of mutual information.
\end{abstract}
\section{Introduction} \label{intro}

Understanding the evolution of many-body systems in generic out-of-equilibrium configurations is a topic of significant interest. If a system is initially prepared in a pure state, it will evolve unitarily and remain in a pure state. However, a finite subregion of a quantum system is expected to thermalize in response to an external disturbance. A useful order parameter to study thermalization in quantum systems is entanglement entropy. To compute this quantity, one can consider dividing the system into two regions: \( A \) and its complement \( A^c \), such  that the Hilbert space factorizes as \( \mathcal{H}_{\text{total}} = \mathcal{H}_A \otimes \mathcal{H}_{A^c} \). Then entanglement entropy between \( A \) and \(A^{c}\) is given by the von Neumann entropy
\begin{equation}
	S_{A}=-\text{tr}_{A} \left[\rho_{A}\log \rho_{A} \right].
\end{equation}
Here, \( \rho_A = \text{tr}_{A^c}[\rho] \) is the reduced density matrix of region \( A \). Because of its necessarily non-local character, entanglement entropy, in principle, can access quantum correlations that are not encoded in observables built out of any collection of local operators.

One of the simplest and most widely studied scenarios is a `quantum quench', where a sudden change in system parameters induces non-equilibrium dynamics. In the Hamiltonian (or Lagrangian) formalism, a quantum quench is typically described by introducing a time-dependent perturbation over the initial Hamiltonian $H_{0}$ (or the Lagrangian $\mathcal{L}_{0}$)
\begin{equation*}
	H(t) = H_{0} + \lambda(t) \delta H_{\Delta},\quad \mathcal{L}(t) = \mathcal{L}_{0} + \lambda(t) \mathcal{O}_{\Delta}, 
\end{equation*}
where $\lambda(t)$ corresponds to a tunable external parameter, and $H_{\Delta}$ $\left(\text{or } \mathcal{L}_{\Delta} \right)$ represents a deformation by an operator of conformal dimension $\Delta$. When the external parameter is spatially homogeneous, the system undergoes a `global quench' -- the type of perturbation we focus on in this paper.

 In the seminal work \cite{Calabrese:2005in}, it was shown that for a subsystem of length $\ell = 2R$ in a $\left(1+1\right)$ dimensional conformal field theory (CFT), entanglement entropy grows linearly in time after an instantaneous quench (i.e. $\lambda(t) \sim \delta(t)$), provided that $\ell \gg \beta$, where $\beta$ is the inverse temperature of the final state
\[\Delta S_{A}(t) = 2\, t\, s_{\text{eq}},\quad t \leq t_{\text{sat}}, \]
where $\Delta S_{A}$ is the difference of entanglement entropy from the ground state, and $s_{\text{eq}}$ is the entanglement entropy of the final state.

In higher dimensions, where analytical tools for computing entanglement entropy are limited, the AdS/CFT correspondence \cite{Maldacena:1997re, Gubser:1998bc, Witten:1998qj} or holography is a useful framework to study CFTs with a dual gravity description. In the holographic setting, a global quench on the field theory side is typically modelled by the formation of a black hole in the bulk. For holographic CFTs the growth of entanglement after an instantaneous global quench is found to exhibit a universal linear regime \cite{Abajo-Arrastia:2010ajo, Albash:2010mv, Hartman:2013qma, Liu:2013iza, Liu:2013qca}
\[\Delta S_{A}(t) = v_{E} s_{\text{eq}} A_{\Sigma} t,\quad t_{\text{loc}} \leq t \leq t_{\text{sat}}, \]
where $v_{E}$, known as the `entanglement velocity', depends on the no. of spacetime dimensions and the parameters of the final state. The quantity $A_{\Sigma}$ is the area of the boundary of the entangling subregion. The time scales $t_{\text{loc}}$ and $t_{\text{sat}}$ correspond to local equilibration and saturation, respectively, with entanglement entropy reaching its steady state value for $t \geq t_{\text{sat}}$. This finding is universal with respect to the shape of the entangling region, and is explained by the `entanglement tsunami' picture \cite{Liu:2013iza, Liu:2013qca}. Also see \cite{Alishahiha:2014cwa, Fonda:2014ula, Alishahiha:2014jxa} for a generalisation of this phenomenon for some non-relativistic holographic spacetimes.

For small subsystem sizes or quenches that are not instantaneous, the behaviour of entanglement entropy follows a different pattern. In this case, since $\ell \ll \beta$ and $t_{\text{sat}} \ll t_{\text{loc}}$, the subsystem never actually enters the regime where the linear growth formula applies. Nonetheless, it has been discovered \cite{Kundu:2016cgh, Lokhande:2017jik, Lokhande:2018nrq} that for small subsystems the growth of entanglement entropy exhibits a different kind of universality -- one that depends on the subsystem shape, but remains independent of the parameters of the final state. This universality persists even for a quench of finite duration \cite{Lokhande:2017jik}.

In this article, we study the thermalization of a small subsystem by analysing the evolution of entanglement entropy in a $d$-dimensional holographic conformal field theory, whose $\left(d+1\right)$-dimensional bulk dual is given by Einstein-Gauss-Bonnet gravity. In this case, the holographic entanglement entropy (HEE) proposal \cite{Ryu:2006bv, Ryu:2006ef, Hubeny:2007xt} must be modified to account for higher derivative corrections in the spacetime metric. We adopt the approach of \cite{Hung:2011xb, deBoer:2011wk}, which relates entanglement entropy to the area of a co-dimension two bulk hypersurface that extremizes the Jacobson-Myers functional \cite{Jacobson:1993xs}. For analytical control, we perform our calculations in the limit where both the subsystem size $\ell$ and the Gauss-Bonnet parameter $\alpha$ are small. We show that while the inclusion of Gauss-Bonnet corrections to Einstein-Hilbert gravity affects the rate of entanglement entropy growth and the time to attain saturation -- the universality observed in \cite{Kundu:2016cgh, Lokhande:2017jik} remains intact.

Entanglement entropy of time-independent excited states perturbatively close to the vacuum obeys a property known as the `first law of entanglement' \cite{Bhattacharya:2012mi, Allahbakhshi:2013rda, Wong:2013gua}. Mathematically this is usually stated as
\begin{equation} \label{eq:first_law}
	T_{A} \Delta S_{A} = \Delta E_{A},
\end{equation}
where $\Delta E_{A}$ is the change in energy within the subsystem $A$, and $T_{E}$ is a quantity known as the `entanglement temperature' which scales as the inverse of the subsystem size $T_{A} \sim \ell^{-1}$ for relativistic CFTs. The first law relationship is strictly valid for small subsystems $\ell \ll \beta$.

For time-dependent excited states formed after a global quench, the relationship \eqref{eq:first_law} is not expected to hold, as such states are not perturbatively close to the vacuum. In \cite{Lokhande:2017jik}, it was shown that even for such states, the quantity
\begin{equation}
	\Upsilon_{A}(t) = \frac{\Delta E_{A}(t)}{T_{A}} - \Delta S_{A}(t),
\end{equation}
always remains positive. This quantity $\Upsilon_{A}(t)$ can be thought of as a time-dependent analogue of relative entropy, making the last equation a generalization of the positivity of relative entropy for time-dependent excited states. In the present article, we also verify this statement for holographic Gauss-Bonnet gravity.

Additionally, we look at the evolution of holographic mutual information 
after a global quench in holographic Gauss-Bonnet gravity. For a boundary subregion $A$ that itself is made by two disjoint regions $A = A_{1} \cup A_{2}$, with $A_{1} \cap A_{2} = \emptyset$, the mutual information is usually defined as
\begin{equation} \label{eq:MI_defn}
	I\left(A_{1} \cup A_{2} \right) = S_{A_{1}} + S_{A_{2}} - S_{A_{1} \cup A_{2}}.
\end{equation}
The advantage of this quantity is that unlike entanglement entropy, mutual information does not suffer from any short-distance divergence. Quantum mechanically $I\left(A_{1} \cup A_{2} \right)$ measures total correlation -- including classical correlation and quantum entanglement between the two subsystems $A_{1}$ and $A_{2}$. The strong subadditivity property of entanglement entropy implies that mutual information is always non-negative: $I\left(A_{1} \cup A_{2} \right) \geq 0$. Using gauge/gravity duality mutual information is studied in \cite{Hubeny:2007re, Headrick:2010zt, Tonni:2010pv}. It is observed that mutual information exhibits a phase transition from positive values to zero as the distance between the two disjoint regions is increased. Time-dependence of mutual information in holographic systems is also studied in \cite{Balasubramanian:2011at, Allais:2011ys, Callan:2012ip, Li:2013sia, Ziogas:2015aja}.

The rest of the article is organized as follows: in section \ref{sec2} we discuss the key features of an asymptotically AdS charged black hole solution in $\left(d+1\right)$-dimensional Einstein-Gauss-Bonnet gravity. We study the growth of entanglement for small subsystems using a perturbation series approach, with results for strip and ball subsystems collected in subsections \ref{subsec 3.2} and \ref{subsec 3.3}. In section \ref{sec4} we identify different regimes during the thermalization process and find out the approximate behaviour of holographic entanglement entropy in these regimes. Section \ref{sec5} is devoted to the analysis of linear response, where we focus on a time-dependent quantity that approaches relative entropy in equilibrium. We study the evolution of mutual information after the global quench in section \ref{sec6}. Finally, we present our conclusions in section \ref{conclusions}.

Before proceeding further, we would like to note some additional references. Quantum quench dynamics in holographic CFTs dual to Einstein-Gauss-Bonnet gravity  has been studied before in \cite{Zeng:2013mca, Li:2013sia, Zeng:2013fsa, Caceres:2013dma, Ghaffarnejad:2018vry, Li:2021rff} using different observables. To the best of our knowledge, the linear response of entanglement entropy for different kinds of quenches, and in the presence of a chemical potential has not been explored. This is the new aspect of the present article.

\section{Charged Vaidya-AdS black brane in Gauss-Bonnet gravity} \label{sec2}

We discuss electrically charged, asymptotically AdS black branes in $\left(d+1\right)$ spacetime dimensions. They are solutions of Einstein-Gauss-Bonnet gravity with a negative cosmological constant, minimally coupled to a $U\left(1\right)$ gauge field. The action of the theory reads
\begin{equation} \label{eq:GB_action}
	I = \frac{1}{16\pi G_{N}} \int_{\mathcal{M}} d^{d+1}x \sqrt{-g} \left(R + \frac{d\left(d+1\right)}{L^2} - 4\pi G_{N} F_{\mu\nu}F^{\mu\nu} + \alpha L_{\mathrm{GB}} \right),
\end{equation}
where $G_{N}$ is the $\left(d+1\right)$ dimensional Newton's constant, $R$ is the Ricci scalar, $F_{\mu\nu} = \partial_{\mu}A_{\nu} - \partial_{\nu}A_{\mu}$ is the $U(1)$ field strength, and $L$ is a length scale that determines the curvature of the asymptotically AdS geometry. The cosmological constant $\Lambda$ is given by $\Lambda = - \frac{d\left(d-1\right)}{2 L^2}$; for notational convenience we henceforth set $L=1$. Finally, the Gauss-Bonnet term is
\begin{equation}
	L_{\mathrm{GB}} = R^2 - 4 R_{\mu\nu}R^{\mu\nu} + R_{\mu\nu\rho\sigma}R^{\mu\nu\rho\sigma}.
\end{equation}
The coefficient $\alpha$ is a constant parameter. It is well-known that the action \eqref{eq:GB_action} admits charged black brane solution for which the metric and non-zero components of gauge field are given by \cite{Cai:2001dz, Astefanesei:2008wz, Anninos:2008sj, Zeng:2013fsa}
\begin{subequations}
\begin{align}
	ds^2 &= \frac{1}{z^2}\left(-F(z) dt^2 + \frac{dz^2}{F(z)} + dx_{i}dx^{i} \right),\quad i = 1, 2, \ldots, d-1, \label{eq:black_metric}\\
	F(z) &= \frac{1}{2\tilde{\alpha}}\left(1 - \sqrt{1 - 4\tilde{\alpha}\left(1 - M z^{d} + Q^2 z^{2d-2} \right) } \right),\quad  \tilde{\alpha} = \left(d-2\right) \left(d-3\right) \alpha, \label{F(z)}\\
	A_{t} &= - \frac{Q}{4\pi\left(d-2\right)}\left(z_{h}^{d-2} - z^{d-2} \right) = \mu \left(1 - \frac{z^{d-2}}{z_{h}^{d-2}} \right).
\end{align}
\end{subequations}
In the above expressions, $z_{h}$ represents the outer event horizon, defined to be the largest positive solution of $F\left(z_{h}\right) = 0$, and $\mu \equiv \displaystyle{\lim_{z \to 0} A_{t}} = -\frac{Q z_{h}^{d-2}}{4\pi\left(d-2\right)}$ is the electrostatic potential, which may be identified with the chemical potential of the dual QFT. $M$ and $Q$ are two conserved quantities which determine the mass $M_0$ and charge $Q_{0}$ of the black brane
\begin{equation}
	M_{0} = \frac{\left(d-1\right) \Omega_{d-1}}{16\pi G_{N}}M,\qquad Q_{0}^2 = \frac{2\pi\left(d-1\right)\left(d-2\right)}{G_{N}}Q^2,
\end{equation}
where $\Omega_{d-1}$ is the volume of the transverse $(d-1)$ dimensional space. It can be easily checked that the black brane metric in equation \eqref{eq:black_metric} asymptotically $\left(z\to 0\right)$ approaches an anti-de Sitter spacetime 
\begin{align}
	ds^2 = \frac{1}{L^2_{\text{eff}}} \left(-dt^2 + L^2_{\text{eff}}\, dx_{i}dx^{i} \right) + \frac{L^2_{\text{eff}}}{z^2}dz^2,
\end{align}
of radius given by
\begin{equation} \label{eq:AdS_Leff}
	L_{\mathrm{eff}}^2 = \frac{2\tilde{\alpha}}{1-\sqrt{1-4\tilde{\alpha}}}.
\end{equation}
For future reference, let us also note the Hawking temperature
\begin{equation}
	T = \frac{d}{\pi z_{h}}\left(1 - \frac{d-2}{d} Q^2 z_{h}^{2d-2} \right).
\end{equation} 

To obtain a Vaidya-type evolving black brane, we change to the ingoing Eddington-Finkelstein coordinate by introducing the null time
\begin{equation} \label{eq:null-time_defn}
	dv = dt - \frac{dz}{F(z)},
\end{equation}
and letting the mass and charge parameters depend on $v$. Thus we arrive at
\begin{subequations}
\begin{align} \label{eq:Vaidya_metric}
	ds^2 &= \frac{1}{z^2}\left(-F(v,z) dv^2 - 2 dv dz + dx^{i}dx_{i} \right),\\
	F(v, z) &= \frac{1}{2 \tilde{\alpha}}\left(1 - \sqrt{1 - 4 \tilde{\alpha} \left(1 - M(v) z^d + Q^2(v) z^{2d-2} \right) } \right).
\end{align}
\end{subequations}

For our purposes, it is convenient to rewrite the Vaidya type solution \eqref{eq:Vaidya_metric} in a different parametrization. Instead of using $M(v)$ and $Q(v)$, we express $F(z)$ in terms of an apparent outer event horizon $z_{h}(v)$ and an auxiliary function $\varepsilon(v)$
\begin{equation} \label{eq:Vaidya_black_fn}
	F(v, z) = \frac{1}{2 \tilde{\alpha}}\left(1 - \sqrt{1 - 4 \tilde{\alpha} \left(1 - \left(\varepsilon(v) \frac{z^d}{z_h(v)^d} - \left(\varepsilon(v)-1\right) \frac{z^{2d-2}}{z_h(v)^{2d-2}} \right) \right) } \right),
\end{equation}
where $\varepsilon(v)$ is related to the original black hole parameters through the relation
\begin{equation} \label{eq:Vaidya_epsilon_defn}
	\varepsilon(v) = 1 + Q(v)^2 z_{h}(v)^{2d-2},
\end{equation}
and at this stage the parameters are arbitrary functions of the null time coordinate $v$. Later, we will choose their functional dependence in different ways-- which will determine the way the global quench is achieved in the dual CFT.

In the time dependent scenario, we may also define a notion of apparent temperature and chemical potential by considering the upgrades \( T \rightarrow T(v) \) and \( \mu \rightarrow \mu(v) \). We define the function \( T(v) \) as follows
 
 \[
 T(v) \equiv -\frac{1}{4\pi} \left. \frac{d F(v, z)}{dz} \right|_{z_{h}(v)} = \frac{2(d - 1) - (d - 2) \varepsilon(v)}{4\pi z_{h}(v)}\,. \label{eq:apparent_temp}
 \]
 Strictly speaking, \( T(v) \) can be associated with the physical temperature only in the limits \( v \to -\infty \) and \( v \to \infty \), which represent the initial and final states, respectively. In between these limits, the system is out of equilibrium, and thermodynamic quantities are ill-defined. Similarly, a time-dependent apparent chemical potential \( \mu(v) \) is defined as 
 \begin{equation}
 	\mu(v) = -\frac{1}{4\pi z_{h}(v)} \frac{\sqrt{\varepsilon(v)-1}}{d-2}.
 \end{equation}
We may even invert these equations to obtain $z_{h}$ and $\varepsilon$ in terms of the CFT parameters
\begin{equation} \label{eq:geometry_intermsof_CFT}
	\begin{split}
		z_{h} &= \frac{-\frac{T}{(d-2)^3} \left(1 - \sqrt{1 + 4 d \left(d-2 \right)^3 \frac{\mu^2}{T^2}}\right)}{8 \pi \mu^2},\\
		\varepsilon &= 2 + \frac{2}{d-2} + \frac{T^2 \left(1 - \sqrt{1 + 4 d (d-2)^3 \frac{\mu^2}{T^2}}\right)}{2 (d-2)^4 \mu^2}.
	\end{split}
\end{equation}
Sometimes it is useful to introduce an effective temperature $T_{\mathrm{eff}}(v)$ given by \cite{Kundu:2016cgh}
\begin{equation} \label{eq:T_eff}
	T_{\mathrm{eff}}(v) = \frac{d}{4\pi z_{h}} = \frac{T}{2}\left(1 + \sqrt{1 + 4 d \left(d-2\right)^3 \frac{\mu^2}{T^2}}  \right),
\end{equation}
which serves as the dominant scale in the theory as it interpolates between $T_{\mathrm{eff}} \propto T$ and $T_{\mathrm{eff}} \propto \mu$ in the limits $\frac{\mu}{T} \ll 1$ and $\frac{\mu}{T} \gg 1$, respectively. Explicitly, we see that
\begin{align*}
	\frac{\mu}{T} \ll 1:\quad T_{\mathrm{eff}}(v) &= T\left(1 + d (d-2)^3 \frac{\mu^2}{T^2} + \mathcal{O}\left(\frac{\mu^4}{T^4} \right) \right),\\
	\frac{\mu}{T} \gg 1:\quad T_{\mathrm{eff}}(v) &= \mu \sqrt{d\left(d-2\right)^3} \left(1 + \frac{1}{2\sqrt{d\left(d-2\right)^3}} \frac{T}{\mu} + \mathcal{O}\left(\frac{T^2}{\mu^2} \right) \right).
\end{align*}

\section{Evolution of holographic entanglement entropy} \label{sec3}

The Ryu-Takayanagi formula \cite{Ryu:2006bv, Ryu:2006ef} associates the entanglement entropy between any subsystem $\left(A\right)$ and its complement in a $d$ dimensional CFT with a co-dimension two minimal area hypersurface $\left(\gamma_{A}\right)$-- which is homologous to the boundary subsystem, in the dual $\left(d+1\right)$ dimensional static, asymptotically AdS solution of Einstein-Hilbert gravity. For time-dependent states, the condition of minimality is replaced by extremality \cite{Hubeny:2007xt}
\begin{equation}
	S_{A} = \frac{1}{4 G_{N}} \int_{\gamma_{A}} \sqrt{h},\quad \partial A \sim \partial \gamma_{A},
\end{equation}
with $h = \mathrm{det}\left(h_{ab}\right)$ being the determinant of the induced metric on $\gamma_{A}$.

In Gauss-Bonnet gravity, higher derivative corrections modify the extremal area functional, necessitating the Jacobson-Myers functional approach to account for these corrections. The corresponding formula is given by \cite{Fursaev:2006ih, deBoer:2011wk, Hung:2011xb, Bhattacharyya:2013jma}
\begin{equation} \label{eq:HEE-GB_formula}
	S_{A} = \frac{1}{4 G_{N}} \int_{\gamma_{A}} \sqrt{h} \left(1 + 2\alpha \mathcal{R}\right) + \frac{1}{2 G_{N}} \int_{\partial\gamma_{A}} \sqrt{\tilde{h}} \left(2\alpha \mathcal{K}\right),\quad \partial A \sim \partial \gamma_{A},
\end{equation}
where $\mathcal{R}$ is the Ricci scalar of the co-dimension two hypersurface $\gamma_{A}$, $\tilde{h}$ is the determinant of the induced metric (of the induced metric $h_{ab}$) on the boundary $\partial\gamma_{A}$, and $\mathcal{K}$ is the extrinsic curvature of $\partial\gamma_{A}$. The additional boundary term in \eqref{eq:HEE-GB_formula} ensures a well-posed variational problem, modifying the UV-divergent contribution without altering the leading-order behaviour of entanglement entropy.

The charged AdS black brane \eqref{eq:black_metric} is asymptotically AdS with AdS radius $L_{\text{eff}}^2$, and boundary coordinates $\tilde{x}^{j} = L_{\text{eff}}\,x^{j}$. We calculate the holographic entanglement entropy associated with two representative boundary subregions which are 
\begin{itemize}

\item A $\left(d-1\right)$ dimensional strip of width $\ell$, specified by
	\begin{equation*}
		A:\quad x_{1} \in \left[-\frac{L_{\text{eff}}\ell}{2}, \frac{L_{\text{eff}}\ell}{2}\right],\quad x_{j} \in \left[0, L_{\text{eff}}\ell_{\perp}\right],\quad j = 2, 3, \ldots, d-1,
	\end{equation*}
	with $\ell_{\perp} \to \infty$. The lengths $\left(\ell, \ell_{\perp}\right)$ are boundary separation in the black brane geometry. The subregion admits translation symmetry in the transverse $\vec{x}_{\perp}$ directions, which should be reflected in the corresponding extremal hypersurface $\gamma_{A}$. Therefore, we choose to parametrize $\gamma_{A}$ using two functions $v = v(z)$, and $x_{1} = x(z)$. Further, the extremal hypersurface is subject to the boundary conditions
	\begin{equation*}
		x(0) = \pm \frac{L_{\text{eff}}\,\ell}{2},\quad v(0) = t.
	\end{equation*}
	Using the spacetime metric written as in equation \eqref{eq:Vaidya_metric}
	\begin{align}
		ds^2 &= \frac{1}{z^2}\left(-F(v,z) dv^2 - 2 dv dz + dx^{i}dx_{i} \right),
	\end{align}
	it is easy to obtain
	\begin{subequations}
		\begin{align}
			\sqrt{h} \mathcal{R} &= \frac{(d-2)(d-3)}{z^{d-1} \sqrt{x'^2 - F(v,z) v'^2 - 2 v'}} + \frac{d}{dz} \left(\frac{2(d-2)}{z^{d-2} \sqrt{x'^2 - F(v,z) v'^2 - 2 v'}} \right),\\
			\sqrt{\tilde{h}} \mathcal{K} &= - \frac{d-2}{z^{d-2} \sqrt{x'^2 - F(v,z) v'^2 - 2 v'}}.
		\end{align}
	\end{subequations}
	Therefore, upon using Stokes' theorem, the contribution from the boundary term is seen to cancel a part of the bulk contribution in \eqref{eq:HEE-GB_formula}. Finally, we derive the following analogue of the area functional for holographic entanglement entropy of a strip
	\begin{align}
		\mathcal{A}(t) &= 2 L_{\text{eff}}^{d-2} \ell_{\perp}^{d-2} \int_{0}^{z_{*}} dz \frac{\sqrt{x'^2 - F(v, z) v'^2 - 2 v'}}{z^{d-1}} \left(1 + \frac{2\left(d-2\right) (d-3) \alpha}{x'^2 - F(v, z) v'^2 - 2 v'} \right), \label{eq:functional-strip}
	\end{align}
	where prime denotes derivative w. r. t. $z$. The limits of integration are taken from the boundary $\left(z = 0\right)$ till a turning point $z = z_{*}$; which is defined to be the point of maximum approach by the HRT hypersurface where $x\left(z_{\ast}\right) = 0$.

\item A $\left(d-1\right)$-dimensional ball of radius $R$, which is defined as
	\begin{equation*}
		A:\quad r^2 \equiv \sum_{i = 1}^{d-1} x_{i}^2 \leq \left(L_{\text{eff}}\,R\right)^2.
	\end{equation*}
	For this geometry, it is convenient to rewrite the spatial part of the boundary metric in equation \eqref{eq:Vaidya_metric} in spherical polar coordinates
	\begin{equation}
		dx_{i} dx^{i} = dr^2 + r^2 d\Omega^2_{d-2},
	\end{equation}
	where $d\Omega^2_{d-2}$ is the metric on the unit $\left(d-2\right)$ dimensional sphere $S^{d-2}$. Once again we make use  of the symmetry of the entangling region to parametrise the hypersurface using two functions $r = r(z), v = v(z)$. In this case the area functional for entanglement entropy takes the following form
	\begin{align}
		\mathcal{A}\left(t\right) &= A_{d-2} \int_{0}^{z_{*}} \frac{dz\,r(z)^{d-2}}{z^{d-1}} \sqrt{r'^2 - F(v, z) v'^2 - 2 v'} \left(1 + \frac{2(d-2)(d-3)\alpha }{\left(r'^2 - F(v, z) v'^2 - 2 v'\right)^2}\right), \label{eq:functional-ball}
	\end{align}
 Where $A_{d-2}=\frac{2\pi^{\frac{d-1}{2}} \left(L_{\text{eff}}\,R\right)^{d-2}}{\Gamma \left(\frac{d-1}{2} \right)}$ is the area of a $\left(d-2 \right)$-dimensional sphere of radius $L_{\text{eff}}\, R$. The extremal hypersurface for the ball subregion must also obey the boundary conditions
	\begin{equation*}
		r(0) = L_{\text{eff}}\, R,\quad v(0) = t.
	\end{equation*}
\end{itemize}

\subsection{Perturbative expansion for small subregion} \label{subsec 3.1}
It is an arduous task to obtain a complete analytical solution of the Euler-Lagrange equations for the extremal hypersurface in its full generality even for Einstein-Hilbert gravity. For small subsystems, there exist perturbative methods which can be utilized to calculate the change in entanglement entropy with respect to the ground state -- which is dual to a pure anti-de Sitter spacetime.

In order to compute the leading order change in entanglement entropy, we proceed in the following way: let us denote by $\mathcal{L}\left[\phi_{A}(z), \lambda\right]$ the Lagrangian functional for the extremal hypersurface. Here, $\phi_{A}(z)$ represents all embedding functions involved in the Lagrangian, and $\lambda$ is a small, dimensionless parameter that acts as a perturbation, i.e. $\lambda<<1$. With this setup, we can expand both $\mathcal{L}$ and $\phi_{A}(z)$ with respect to $\lambda$ as follows
\begin{equation} \label{eq:embedding_equation}
	\begin{split}
	\mathcal{L}\left[\phi_{A}(z), \lambda \right] &= \mathcal{L}^{(0)}\left[\phi_{A}(z) \right] + \lambda\mathcal{L}^{(1)}\left[\phi_{A}(z) \right] + \mathcal{O}(\lambda^2),\\
	\phi_{A}(z) &= \phi_{A}^{(0)}(z) + \lambda\phi_{A}^{(1)}(z) + \mathcal{O}(\lambda^2).
	\end{split}
\end{equation}  
In principle, the embedding functions $\phi^{(n)}_{A}(z)$ can be determined by solving the equations of motion derived from the functional at every order. However, because these equations are generally non-linear, finding solutions can be very challenging or even impossible. Notwithstanding this difficulty, if we focus our attention only at the leading order $\mathcal{O}\left(\lambda\right)$, we find that the on-shell area functional is
\begin{equation}
	\begin{split}
	\mathcal{A} \left[\phi_{A}(z)\right] = \int &dz \, \mathcal{L}^{(0)}\left[\phi_{A}^{(0)}(z)\right] + \lambda\int dz \, \mathcal{L}^{(1)}\left[\phi_{A}^{(0)}(z)\right] \\
	&+ \lambda \int dz\, \phi^{(1)}_{A}(z) \left[-\frac{d^2}{dz^2}\left(\frac{\partial\mathcal{L}^{(0)}}{\partial\phi^{''}_{A}(z)}\right) + \frac{d}{dz} \frac{\partial\mathcal{L}^{(0)}}{\partial\phi^{'}_{A}(z)}-\frac{\partial\mathcal{L}^{(0)}}{\partial\phi_{A}(z)}\right]_{\phi_{A}^{(0)}} + \mathcal{O}\left(\lambda^2\right)
	\end{split}
\end{equation}
The term inside square brackets in the second line above is the Euler-Lagrange equation for the embedding function in the ground state. Therefore, it evaluates to zero on-shell; and we are left with
\begin{equation}
	\mathcal{A}\left[\phi_{A}(z)\right] = \int dz \, \mathcal{L}^{(0)}\left[\phi_{A}^{(0)}(z)\right] + \lambda\int dz \, \mathcal{L}^{(1)}\left[\phi_{A}^{(0)}(z)\right]+\mathcal{O}(\lambda^2) 
\end{equation}
The fortunate aspect of this result is that we require only the zeroth order embedding functions $\phi_{A}^{(0)}(z)$, for which analytical solutions are easier to obtain, to calculate the leading order change in entanglement entropy. In the present case of interest, the perturbation parameter $\lambda$ is taken to be $\lambda = \frac{z_{\ast}^{d}}{z_{h}^{d}} \ll 1$. From the perspective of the boundary CFT, the turning point $z_{\ast} \sim \ell$, the characteristic length of the subregion. While $z_{h}$ being the horizon is related to the inverse of the effective temperature $z_{h} \sim T_{\text{eff}}^{-1}$ according to equation \eqref{eq:T_eff}. Therefore, the perturbative expansion in the CFT is carried out in powers of $T_{\text{eff}}\ell$.

\subsection{Holographic entanglement entropy of the strip} \label{subsec 3.2}

For the strip subregion we have to find out a $\left(d-1\right)$-dimensional bulk hypersurface which extremizes the modified area functional in equation \eqref{eq:functional-strip}. To simplify notation, we describe the calculation in five bulk spacetime dimensions without loss of generality. The results for six and seven spacetime dimensions are given in appendix \ref{appendix:other_d}.

We use the small subregion size approximation described in the previous section, and assume $\lambda = \frac{z_{*}}{z_{h}} \ll 1$. Therefore, the blackening function $F(v, z)$ can be approximately written as
\begin{equation}\label{F fn quench}
	F(v, z) = \frac{1 - \sqrt{1 - 8\alpha}}{4\alpha} - \frac{g(v) \varepsilon \lambda^4 z^4}{\sqrt{1-8\alpha}} + \mathcal{O}\left(\lambda^6\right) ,
\end{equation}
where the time-dependence is now encoded within the function $g(v)$. In writing the above, we have rescaled all the coordinates by the turning-point: $\{v, z, x^{i}\} \to \frac{1}{z_{*}}\{v, z, x^{i}\}$. We also need the embedding functions $\{v(z), x(z)\}$ for the ground state $\left(\lambda = 0\right)$, which is the five dimensional anti-de Sitter spacetime
\begin{equation*}
	ds^2 = \frac{1}{z^2}\left(-\frac{1 - \sqrt{1 - 8\alpha}}{4\alpha} dv^2 - 2 dv dz + dx^{i}dx_{i} \right).
\end{equation*}
Since this is a static spacetime, the extremal hypersurface lies on a constant time slice. Integrating equation \eqref{eq:null-time_defn} for constant $F(v,z)$, it is easy to obtain
\begin{equation} \label{eq:v_soln_strip}
	v = t - \frac{z}{\frac{1 - \sqrt{1 - 8\alpha}}{4\alpha}}.
\end{equation}
Doing a similar expansion of equation \eqref{eq:functional-strip}, and substituting the above expression for $v(t, z)$ we obtain the Lagrangian functional at the zeroth order
\begin{align}
	\mathcal{L}^{(0)} &= \frac{\sqrt{2 x'(z)^2 + 1 + \sqrt{1-8 \alpha }}}{\sqrt{2} z^3} \left(1 + \mathsf{L}_{\text{strip}} \left[x'(z)\right] \right), \label{eq:L0_strip_level1} \\
	\mathsf{L}_{\text{strip}} &= \frac{x'(z)^4 + \left(1 + 4\alpha \right) x'(z)^2 + 2 \alpha \left(2 - \sqrt{1 - 8\alpha}\right)}{x'(z)^4 + x'(z)^2 + 2\alpha}
\end{align}
The zeroth order Lagrangian functional in equation \eqref{eq:L0_strip_level1} does not have any explicit dependence on $x(z)$ and $x''(z)$. Therefore, it admits a first integral of motion
\begin{equation}
	\frac{\partial \mathcal{L}^{(0)}}{\partial x'} = \mathrm{constant}.
\end{equation}
Solving the equation of $x(z)$ is highly non-trivial. Hence, we make another approximation; we consider the Gauss-Bonnet coupling to be very small, i.e. $\alpha \ll 1$, and further expand the equations in powers of $\alpha$ \footnote{$\alpha$ has dimension $\left[\text{length}\right]^2$, and the actual perturbation is carried out in powers of $\frac{\alpha}{L^2}$. Recall that we have set $L=1$ in the beginning.}
\begin{equation} \label{eq:eom_perturbative_strip}
	\frac{x'(z)}{z^3 \sqrt{x'(z)^2+1}} - \frac{3 \alpha  x'(z)}{z^3 \left(x'(z)^2 + 1\right)^{3/2}} = \mathrm{constant}.
\end{equation}
The constant on the r.h.s. can be fixed by noting that $x'(z) \to \infty$ as $z \to 1$ \footnote{We are using dimensionless coordinates, this relationship is really $\left.\frac{dx}{dz}\right|_{z=z_{*}} = \infty$.} Equation \eqref{eq:eom_perturbative_strip} can be solved to obtain a perturbative solution for the hypersurface $x(z)$ in the static, pure anti-de Sitter spacetime at $\mathcal{O}\left(\alpha\right)$
\begin{align}
	x(z) &= \frac{L_{\text{eff}}\,\ell}{2} - \frac{1}{4} z^4\, {}_2F_1 \left[\frac{1}{2}, \frac{2}{3}, \frac{5}{3}, z^6 \right] \left(1 + 3\alpha \right) + \mathcal{O}\left(\alpha^2\right),\\
	&\simeq \left(1 - \alpha \right) \frac{\ell}{2} - \frac{1}{4} z^4\, {}_2F_1 \left[\frac{1}{2}, \frac{2}{3}, \frac{5}{3}, z^6 \right] \left(1 + 3\alpha \right) + \mathcal{O}\left(\alpha^2\right). \label{eq:x_soln_strip}
\end{align}
While this was obtained in units where $z_{*} = 1$, we may easily obtain the relation between the turning point $z_{*}$ and the strip-width $\ell$ using
\begin{align*}
	x\left(z_{*}\right) &= \left(1-\alpha\right)\frac{\ell}{2} - \frac{z_{*}}{4}\, {}_2F_1 \left[\frac{1}{2}, \frac{2}{3}, \frac{5}{3}, 1 \right] \left(1 + 3\alpha \right) = 0. 
\end{align*}
Using $\frac{1}{4}\, {}_2F_1 \left[\frac{1}{2}, \frac{2}{3}, \frac{5}{3}, 1 \right] = \frac{\sqrt{\pi}\, \Gamma\left(\frac{2}{3}\right)}{\Gamma\left(\frac{1}{6}\right)}$, we obtain
\begin{equation} \label{eq:turnpt_strip_instant}
	z_{*} = \frac{\ell}{2} \frac{\Gamma\left(\frac{1}{6}\right)}{\sqrt{\pi}\, \Gamma\left(\frac{2}{3}\right)} \left(1 - 4\alpha\right) + \mathcal{O}\left(\alpha^2\right).
\end{equation}

We use this solution to calculate the leading order change in holographic entanglement entropy by subtracting the zeroth order terms
\begin{equation}
	\label{entropy change 0}
	\Delta S_{A}(t) = \frac{\Delta \mathcal{A}(t)}{4 G_{N}} = \frac{L_{\text{eff}}^2 \ell_{\perp}^2 \lambda^4}{2 G_{N}} \int_{0}^{1} dz \, \left. \mathcal{L}^{(1)}\left[\phi_{A}^{(0)}(z)\right]\right|_{\phi_{A}^{(0)}(z) = x(z)},
\end{equation}
where the on-shell first order change of the Lagrangian up to $\mathcal{O}\left(\alpha^2\right)$, upon using equations \eqref{eq:v_soln_strip} and \eqref{eq:x_soln_strip} can be written as
\begin{equation} \label{eq:L1_final}
	\mathcal{L}^{(1)}\left[x(z)\right] = \frac{z \sqrt{1-z^6}}{2} \left(1 - 3\alpha \right) \varepsilon g \left(t - \frac{z}{\frac{1 - \sqrt{1 - 8\alpha}}{4\alpha}} \right).
\end{equation}
Next, we evaluate the integral by considering different quench profiles 
$g(v)$, obtaining explicit results for the entanglement entropy change.

\subsubsection{Instantaneous quench}

The first example of our choice is an instantaneous quench: where the function $g(v(z))$ is represented by a step function $\Theta\left(v(z)\right)$; i.e. the blackening function $F(v, z)$ takes the form \[ F(v, z) = \frac{1}{2 \tilde{\alpha}}\left(1 - \sqrt{1 - 4 \tilde{\alpha} \left(1 - \Theta(v) \left(\varepsilon \frac{z^d}{z_h^d} - \left(\varepsilon-1\right) \frac{z^{2d-2}}{z_h^{2d-2}} \right) \right) } \right). \] Holographically, an instantaneous quench corresponds to the collapse of an infinitesimally thin shell of null dust. While local operators thermalize immediately, non-local observables such as entanglement entropy require finite time to equilibrate. This makes entanglement entropy a valuable probe of the full thermalization process.

\begin{figure}[t]
	\centering
	\begin{subfigure}[b]{0.48\textwidth}
		\centering  
		\includegraphics[width=\textwidth]{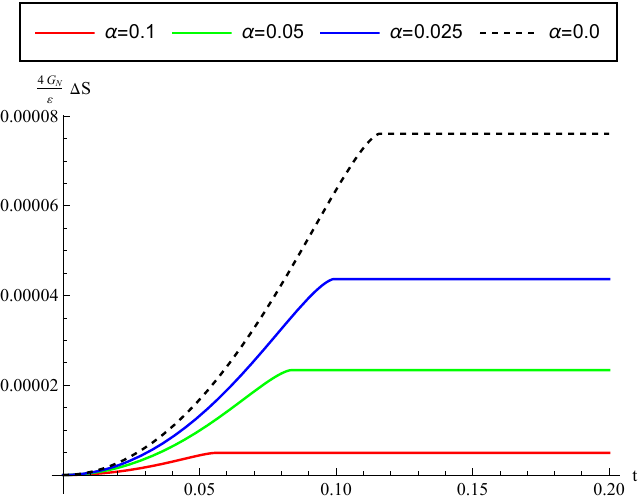}
		\caption{}
		\label{fig:HEE_evolution_instant}
	\end{subfigure}
	\hfill
	\begin{subfigure}[b]{0.48\textwidth}
		\centering  
		\includegraphics[width=\textwidth]{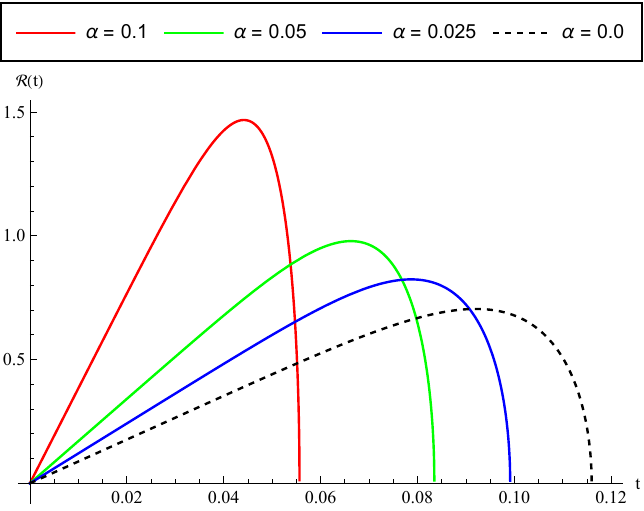}
		\caption{}
		\label{fig:Rate_instant}
	\end{subfigure}
	\caption{Evolution of holographic entanglement entropy for a strip in five-dimensional Gauss-Bonnet gravity, we set $\frac{\ell}{z_{h}}=0.1$ for small subregion, and   $\frac{\ell^2_{\perp}}{G_{N}}=1$ for simplicity. In (a) we show the growth of HEE after an instantaneous global quench for different values of the Gauss-Bonnet coupling $\alpha$. In (b) we plot the instantaneous growth-rate $\mathcal{R}(t)$, we observe that the maximum growth-rate in sensitive to $\alpha$ and may exceed $1$.}
	\label{fig:Instantaneous1}
\end{figure}

To calculate the change in entanglement entropy, we substitute the value of $\mathcal{L}^{(1)}[x(z)]$ from equation \eqref{eq:L1_final} in equation \eqref{entropy change 0} to obtain
\begin{equation}\label{Entropy change}
	\begin{split}
		\Delta S_{A}(t) &= \frac{L_{\text{eff}}^2\ell_{\perp}^{2}\lambda^4\varepsilon }{4G_{N}} \int_{0}^{1} dz\ {z \sqrt{1-z^6}} \left(1-3\alpha \right) \Theta \left(t - \frac{z}{\frac{1 - \sqrt{1 - 8\alpha}}{4\alpha}} \right),\\
		&\simeq \frac{\ell_{\perp}^{2}\lambda^4\varepsilon }{4G_{N}} \int_{0}^{1} dz\ {z \sqrt{1-z^6}} \left(1-5\alpha \right) \Theta \left(t - \left(1 - 2 \alpha \right)z \right).
	\end{split}
\end{equation}
To evaluate this integral, we may introduce a new variable defined as \( \zeta = t - \left(1 - 2 \alpha \right)z\). With this substitution, the integral in equation \eqref{Entropy change} transforms to
\begin{equation}
		I=\int_{t-1+2\alpha}^{t} \frac{d\zeta}{1-2\alpha}\ \Theta(\zeta) \left(\frac{t-\zeta}{1-2\alpha}\right) \sqrt{1- \left(\frac{t-\zeta}{1-2\alpha}\right)^6} \left(1-5\alpha\right).
\end{equation}
We evaluate this by considering three different domains in the time axis, namely: (i) $t<0$, (ii) $0<t<\left(1-2\alpha\right)$, and (iii) $t>\left(1-2\alpha\right)$:
\begin{enumerate}[(i)]
	\item Since both the limits are negative and $\Theta(\zeta<0)=0$, therefore
	\begin{equation}
		I=0.
	\end{equation}
	\item The lower limit is negative so we can replace it by zero. Reverting back to the original variable $z$ we obtain
	\begin{equation}
		\begin{split}
			I &=\int_{0}^{\frac{t}{(1-2\alpha)}} dz\ z \sqrt{1 - z^6} \left(1 - 3\alpha \right),\\
			&= \frac{t^2}{2}\ {_2F_1}\left(-\frac{1}{2},\frac{1}{3},\frac{4}{3},\frac{t^6}{(1-2\alpha)^6}\right) \frac{\left(1-5\alpha\right)}{\left(1-2\alpha\right)^2}
		\end{split}
	\end{equation}
	\item In this interval $\Theta(\zeta>0)=1$ and the integration simplifies to
	\begin{equation}
		I=\int_{0}^{1}dz \ {z \sqrt{1-z^6}}(1-5\alpha) = \frac{\sqrt{\pi}}{10}\frac{\Gamma\left(\frac{1}{3}\right)}{\Gamma\left(\frac{5}{6}\right)}\left(1-5\alpha \right).
	\end{equation}
The last expression is independent of time. Hence entanglement entropy saturates at the time given by $t_{\text{sat}}=1-2\alpha$. Restoring the original coordinates we may write \[t_{\text{sat}} = \left(1 - 2\alpha \right) z_{*}, \] where the turning point $z_{*}$ is given in equation \eqref{eq:turnpt_strip_instant}. As \(\alpha \to 0\), these results align perfectly with the conclusions of \cite{Kundu:2016cgh} in Vaidya-AdS$_5$ black hole solution of two-derivative gravity.

The saturation time $t_{\text{sat}}$ is independent of both the temperature \(T\) and the chemical potential \(\mu\). This result is unsurprising because $t_{\text{sat}}$ is derived using the zeroth-order embedding, which lacks any dependence on the specific properties of the state. The saturation time decreases as the value of \(\alpha\) increases, which can be observed from the plots in Fig. \eqref{fig:HEE_evolution_instant}. However, as mentioned in \cite{Kundu:2016cgh}, the saturation time for small subsystem is influenced by $\frac{\mu}{T}$ after accounting for the first-order correction to the Ryu-Takayanagi hypersurface.
\end{enumerate}

Altogether, the leading order correction to holographic entanglement entropy can be expressed in the following form
\begin{equation} \label{Change in entropy Delta S_A}
	\Delta S_{A}(t) = \Delta S_{\text{eq}}\{\left[\Theta(t) - \Theta\left(t - t_{\text{sat}} \right) \right] \mathcal{F}\left(\frac{t}{t_{\text{sat}}} \right) + \Theta \left(t - t_{\text{sat}} \right)\},
\end{equation}
where $\Delta S_{\text{eq}}$ is the equilibrium value of the entropy
\begin{equation}
	\Delta S_{\text{eq}} = \frac{\varepsilon \ell_{\perp}^{2}}{4G_{N}} \frac{z_{*}^2}{z_{H}^4} \frac{\sqrt{\pi}}{10} \frac{\Gamma \left(\frac{1}{3} \right)}{\Gamma \left(\frac{5}{6} \right)}\left(1-5\alpha \right),
\end{equation}
and $\mathcal{F}$ is given by
\begin{equation}\label{exp of F(t)}
	\mathcal{F}(x) = \frac{5\,\Gamma \left(\frac{5}{6} \right) }{\sqrt{\pi}\ \Gamma \left(\frac{1}{3} \right)}\ x^2\ {_2F_1} \left(-\frac{1}{2}, \frac{1}{3}, \frac{4}{3}, x^6 \right),\quad x = \frac{t}{t_{\text{sat}}}.
\end{equation}
By construction the function \( \mathcal{F}(x) \) takes on the extreme values
\begin{equation}
	\mathcal{F}(0) = 0,\quad \text{and}\quad \mathcal{F}(1) = \frac{5\, \Gamma\left(\frac{5}{6}\right)}{\sqrt{\pi}\,\Gamma\left(\frac{1}{3} \right)}\ {_2F_1}\left(-\frac{1}{2},\frac{1}{3},\frac{4}{3}, 1\right) = 1. 
\end{equation}
Hence its average rate of change over this interval is $\left\langle \frac{d\mathcal{F}(x)}{dx} \right\rangle = 1$. We can now determine the instantaneous rate of entanglement growth\footnote{Here, $s_{\text{eq}} = \Delta S_{\text{eq}} / V_A$ denotes the equilibrium entropy density of the system after the saturation time. $V_A$ is the volume of the entangling subregion $A$, i.e. $V_{A} = \ell \times \ell_{\perp}^{2}$.} following \cite{Kundu:2016cgh}
\begin{equation} \label{Rate eqn}
	\mathcal{R}(t) \equiv \frac{1}{2 s_{\text{eq}}\, \ell_{\perp}^2}\frac{d(\Delta S_{A})}{dt} = \frac{\ell}{2 t_{\text{sat}}} \frac{d\mathcal{F}}{dx} = \frac{\ell}{t_{\text{sat}}} \frac{5 \Gamma\left(\frac{5}{6}\right)}{\sqrt{\pi}\,\Gamma\left(\frac{1}{3}\right)} \frac{t}{t_{\text{sat}}} \sqrt{1 - \left(\frac{t}{t_{\text{sat}}}\right)^6}.
\end{equation}
The time-averaged entanglement velocity is similarly defined as
\begin{equation}
	v_{E}^{\text{avg}} = \langle \mathcal{R}(t) \rangle = \frac{\ell}{2 t_{\text{sat}}} \left\langle \frac{d\mathcal{F}(x)}{dx} \right\rangle = \frac{\ell}{2 t_{\text{sat}}} = \frac{\sqrt{\pi} \Gamma\left(\frac{2}{3}\right)}{\Gamma\left(\frac{1}{6}\right)} \left(1 + 6\alpha \right) + \mathcal{O}\left(\alpha^2\right).
\end{equation}
To obtain the last expression, we used equation \eqref{eq:turnpt_strip_instant} to relate the turning point $z_{*}$ with the strip-width $\ell$. We note that $v_{E}^{\text{avg}} < 1$ courtesy of the bound $\alpha \leq \frac{1}{8}$. This is commensurate with our expectation that in a relativistic theory, entanglement entropy cannot grow faster than the speed of light.
\begin{figure}[t]
	\centering 
	\includegraphics[width =0.5\textwidth]{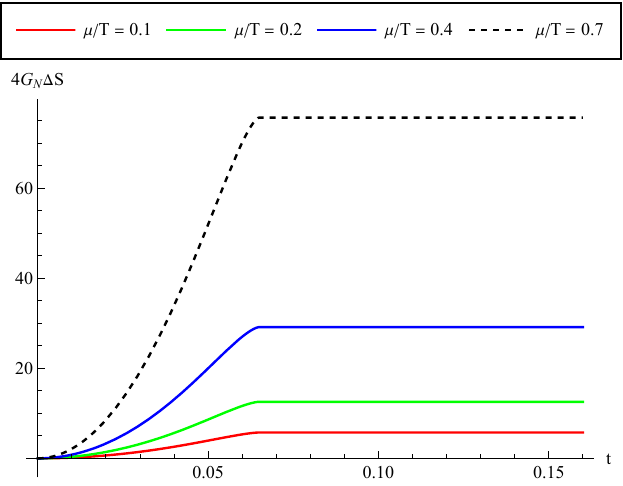} 
	\caption{Evolution of entanglement entropy for a strip after an instantaneous global quench in Gauss-Bonnet gravity. The Gauss-Bonnet coupling is fixed at $\alpha = 0.1$, while the chemical potential-to-temperature ratio $\frac{\mu}{T}$ is varied. Higher $\frac{\mu}{T}$ leads to a greater equilibrium entropy, but saturation time remains unchanged.} 
	\label{HEE Evolution instant wrto CFT} 
\end{figure}

Before diving into a detailed analysis of the different regimes of equation \eqref{Change in entropy Delta S_A}, let's first offer a brief commentary on some general aspects. In Figure \ref{fig:HEE_evolution_instant} we show the evolution of holographic entanglement entropy for different permissible values of the Gauss-Bonnet coupling ($\alpha$) while keeping the charge parameter $\varepsilon$ constant. We observe that larger Gauss-Bonnet coupling leads to shorter thermalization time, although it negatively affects the change in entanglement. This implies that the Gauss-Bonnet coupling accelerates thermalization. In Figure \ref{fig:Rate_instant} we plot the instantaneous rate of growth \eqref{Rate eqn} as a function of time $t$. For an allowed range of $\alpha$ values, it increases with $\alpha$, and its maximum value may even exceed the speed of light. There is no violation of causality whatsoever as we have already checked that the average rate of growth $v_{E}^{\text{avg}}$ always remains below $1$. Nonetheless, this situation is markedly different from the case of holographic theories dual to Einstein-Hilbert gravity, where $\mathsf{max}\left[\mathcal{R}(t)\right]$ was observed to exceed 1 only in two dimensions \cite{Kundu:2016cgh}. Overall, the observed behaviour of the entanglement entropy is qualitatively similar to the  results presented in \citeonline{Kundu:2016cgh, Zeng:2013fsa}.

We also examine the dependence of entanglement growth on the ration $\frac{\mu}{T}$ in the dual CFT, as shown in figure \ref{HEE Evolution instant wrto CFT}. Varying $\frac{\mu}{T}$ changes the effective temperature $T_{\text{eff}}$ in equation \eqref{eq:T_eff}. For these plots, we keep the subsystem size $\ell$ fixed. We find that while the equilibrium value of entanglement entropy increases with $\frac{\mu}{T}$, the saturation time remains unchanged.

\subsubsection{Linear quench}

Let us now examine an example of a finite quench which lasts for a duration \( t_q \). We focus on a family of quenches characterized by power-law behaviour, where the function $g(v(z))$ takes the following form \cite{Lokhande:2017jik}
\begin{equation} \label{Power-Law quench}
	\begin{split}
		&g(v(z)) = v^p(z) \left[\Theta \left(v(z) \right) - \Theta \left(v(z)-t_q \right) \right] + t_{q}^{p}\, \Theta \left(v(z)-t_q \right),\\
		\text{with}\quad &v = t - \left(1 - 2\alpha\right)z + \mathcal{O}\left(\alpha^2 \right).
	\end{split}
\end{equation}
The family of power-law quenches defined above for \(p \in \mathbb{Z}\) is sufficiently general to represent any quench that is analytic within the interval \(t \in (0, t_q)\). For simplicity, we focus on the case $p=1$ in this article, corresponding to a linear quench. The results can be generalized for any $p \in \mathbb{Z}$ without much difficulty.

Again, we can calculate the change in holographic entanglement entropy using the equations \eqref{entropy change 0} and \eqref{eq:L1_final}, with the function \( g(v(z)) \) defined above in equation \eqref{Power-Law quench}. Fortunately, even for a linear quench profile the integrations can be carried out analytically. There are two distinct scenarios one should consider: I. \( t_q < (1-2\alpha) z_{\ast} \), and II. \( t_{q}>(1-2\alpha ) z_{\ast}\). In both cases, the saturation time is given by \( t_{\text{sat}} = t_q + (1-2\alpha) z_{\ast} \), and the evolution can be divided into different intervals for analysis, as depicted in table \ref{table:quenches}.
\begin{table}[h!]
	\centering
	\renewcommand{\arraystretch}{1.5}
	\resizebox{\textwidth}{!}{%
		\begin{tabular}{|c|c|c|c|c|c|}
			\hline
			{Regime} & {Pre-quench} & {Initial} & {Intermediate} & {Final} & {Post-saturation} \\ \hline
			{Case I :} (\(t_q < \left(1-2\alpha \right)z_{\ast}\)) & \(t < 0\)         & \(0 < t < t_q\)   & \(t_q < t < \left(1-2\alpha \right)z_{\ast}\)       & \( \left(1-2\alpha \right) z_{\ast} < t < t_{\text{sat}}\)  & \(t > t_{\text{sat}}\) \\ \hline
			{Case II :} (\(\left(1-2\alpha \right)z_{\ast} < t_q\)) & \(t < 0\)         & \(0 < t < \left(1-2\alpha \right)z_{\ast}\)   & \(\left(1-2\alpha\right)z_{\ast} < t < t_q\)       & \(t_q < t < t_{\text{sat}}\)  & \(t > t_{\text{sat}}\) \\ \hline
		\end{tabular}%
	}
	\caption{Time regimes for the evolution of entanglement entropy after a finite-duration quench.}
	\label{table:quenches}
\end{table}

\begin{figure}[t]
	\centering
	\begin{subfigure}{0.32\textwidth}
		\includegraphics[width=\textwidth]{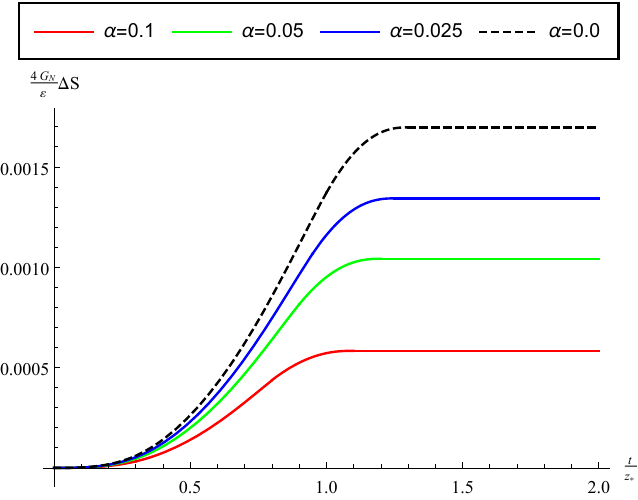}
		\caption{}
	\end{subfigure}
	\hfill
	\begin{subfigure}{0.32\textwidth}
		\includegraphics[width=\textwidth]{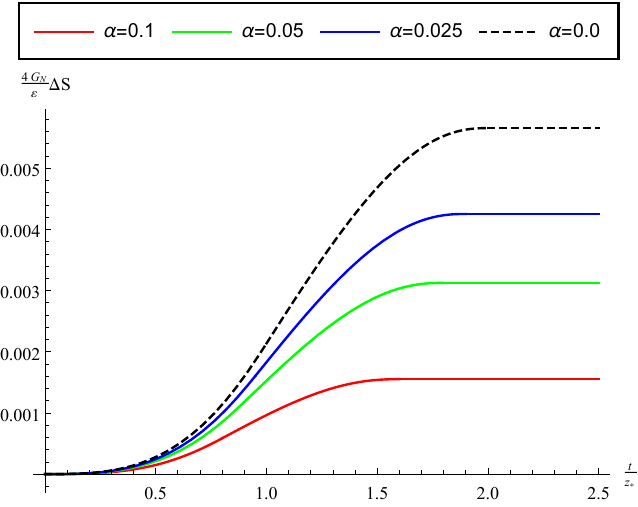}
		\caption{}
	\end{subfigure}
	\begin{subfigure}{0.32\textwidth} 
		\includegraphics[width=\textwidth]{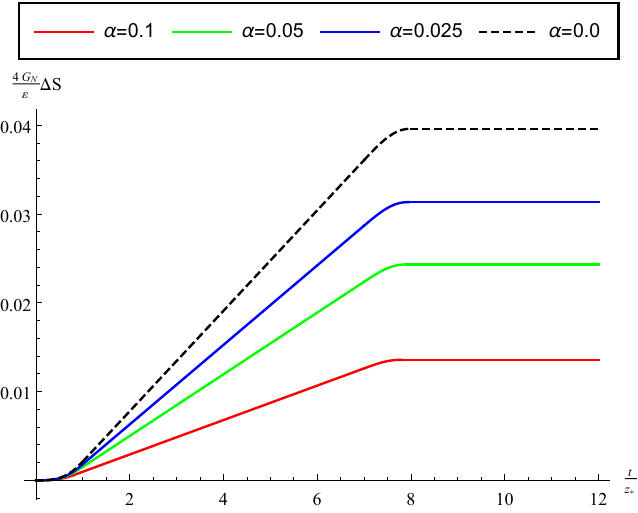}
		\caption{}
	\end{subfigure}
	\hfill
	\begin{subfigure}[b]{0.32\textwidth} 
		\includegraphics[width=\textwidth]{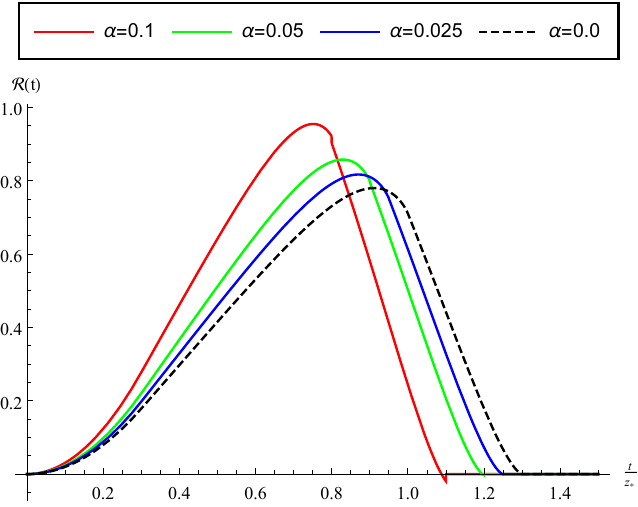}
		\caption{}
	\end{subfigure}
	\begin{subfigure}[b]{0.32\textwidth}
		\includegraphics[width=\textwidth]{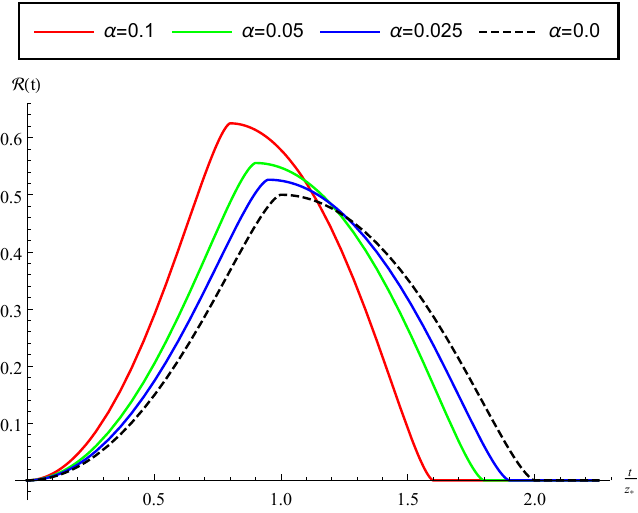}
		\caption{}
	\end{subfigure}
	\hfill
	\begin{subfigure}[b]{0.32\textwidth}  
		\includegraphics[width=\textwidth]{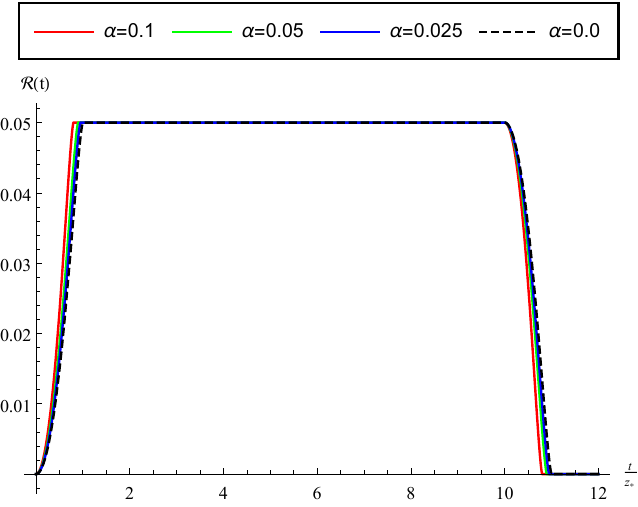}
		\caption{}
	\end{subfigure}
	\caption{Evolution of entanglement entropy $\Delta S(t)$ and its instantaneous growth rate $\mathcal{R}(t)$ for a strip after a linearly driven quench with $p=1$. Different values of Gauss-Bonnet coupling ($\alpha$) are depicted within legend, and we have chosen values $t_q={(0.3, 1-2\alpha, 7)}$for the quench duration in ascending order from left to right.}
	\label{fig:power law all plot}
\end{figure}

\begin{figure}[t]
	\centering
	\begin{subfigure}{0.32\textwidth}
		\includegraphics[width=\textwidth]{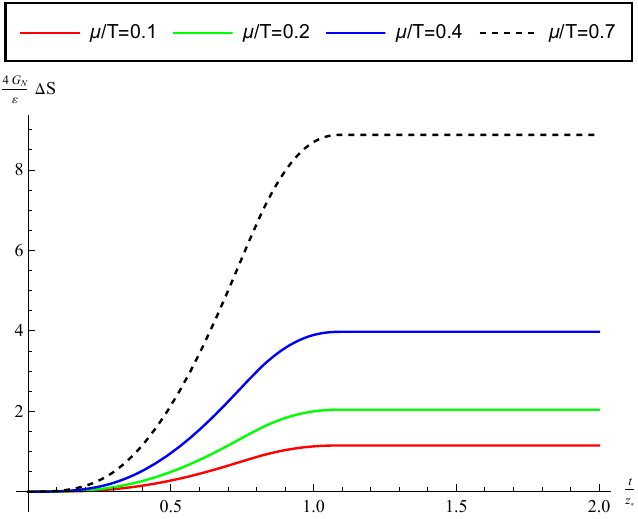}
		\caption{}
	\end{subfigure}
	\hfill
	\begin{subfigure}{0.32\textwidth}
		\includegraphics[width=\textwidth]{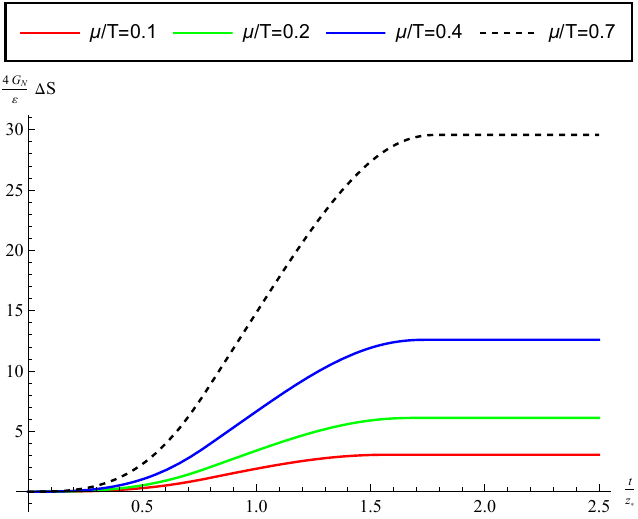}
		\caption{}
	\end{subfigure}
	\begin{subfigure}{0.32\textwidth} 
		\includegraphics[width=\textwidth]{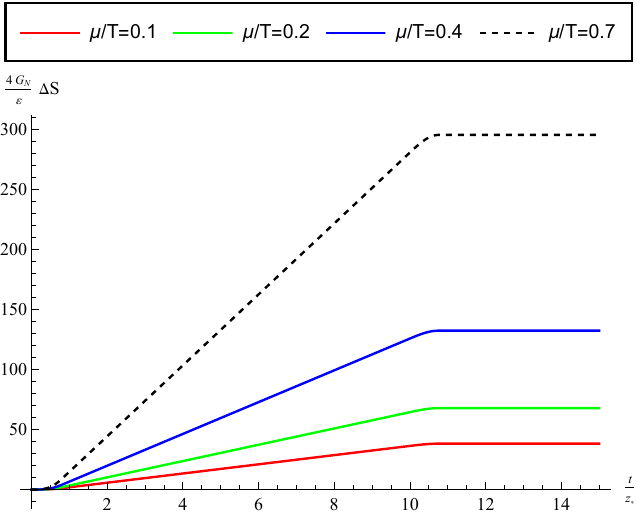}
		\caption{}
	\end{subfigure}
	\caption{Entanglement entropy evolution, $\Delta S_A(t)$, for a strip following a linear quench is examined for different values of $\mu/T$, while holding the Gauss-Bonnet coupling fixed at $\alpha=0.1$. The quench duration $t_q$ is varies across the panels in ascending order from left to right.}
	\label{HEE evolution wrto CFT parameter linaer}
\end{figure}

The pre-quench $\left(t<0\right)$ and post-saturation $\left(t>t_{\text{sat}}\right)$ regions are equilibrium configurations. The initial, intermediate, and final regimes are generally time-dependent. Since the final expressions can be lengthy, let us introduce two indefinite integrals to ease the notation
\begin{subequations}
\begin{align}
	\mathcal{I_{A}}(t,z) &= \frac{\ell_{\perp}^2 z_{*}^2\varepsilon}{4G_{N}z_{H}^4} \int dz\ z\sqrt{1-\frac{z^6}{z_{*}^6}}\ \left(t - \left(1 - 2\alpha \right)z \right) \left(1 - 5\alpha \right), \label{eq:strip_indef_int_A} \\
	\mathcal{I_{B}}(t_q,z) &= \frac{\ell_{\perp}^2 z_{*}^2\varepsilon}{4G_{N}z_{H}^4} \int dz\ z\sqrt{1-\frac{z^6}{z_{*}^6}}\ t_{q} \left(1 - 5\alpha \right). \label{eq:strip_indef_int_B}
\end{align}
\end{subequations}
In terms of these integrals, the change in entanglement entropy \( \Delta S_A(t) \) for the case of a strip can be expressed as follows
\begin{equation}\label{integral for t less}
\Delta S_A^{(I)}(t) = 
\begin{cases}
	0, & t < 0, \\
	\mathcal{I_{A}}(t,z)  |^{\frac{t}{(1-2\alpha)}}_{0}, & 0 < t < t_q, \\
	\mathcal{I_{B}}(t_q,z)  |^{\frac{t-t_q}{(1-2\alpha)}}_{0} +   \mathcal{I_{A}}(t,z)  |^{\frac{t}{(1-2\alpha)}}_{\frac{t-t_q}{(1-2\alpha)}}, & t_q < t < (1-2\alpha)z_{\ast}, \\
	\mathcal{I_{B}}(t_q,z)  |^{\frac{t-t_q}{(1-2\alpha)}}_{0} +   \mathcal{I_{A}}(t,z)  |^{z_{\ast}}_{\frac{t-t_q}{(1-2\alpha)}}, & \left(1-2\alpha\right)z_{\ast} < t < t_{\text{sat}}, \\
	\mathcal{I_{B}}(t_q,z)  |^{z_{\ast}}_{0}, & t > t_{\text{sat}}.
\end{cases}
\end{equation}
and 
\begin{equation}\label{integral for t greater}
\Delta S_A^{(II)}(t) = 
\begin{cases}
	0, & t < 0, \\
	\left. \mathcal{I_{A}}(t,z) \right|^{\frac{t}{(1-2\alpha)}}_{0}, & 0 < t < (1-2\alpha)z_{\ast}, \\
	\left. \mathcal{I_{A}}(t,z) \right|^{z_{\ast}}_{0}, & (1-2\alpha)z_{\ast} < t < t_q, \\
	\left. \mathcal{I_{B}}(t_q,z) \right|^{\frac{t-t_q}{(1-2\alpha)}}_{0} + \left. \mathcal{I_{A}}(t,z) \right|^{1}_{\frac{t-t_q}{(1-2\alpha)}}, & t_q< t < t_{\text{sat}}, \\
	\left. \mathcal{I_{B}}(t_q,z) \right|^{z_{\ast}}_{0}, & t > t_{\text{sat}}.
\end{cases}
\end{equation}
The results can be understood graphically from the plots in figure \ref{fig:power law all plot}, where we show the evolution of entanglement entropy and its rate of growth for some representative cases of the ratio $\frac{t_{q}}{\left(1-2\alpha\right) z_{\ast}} = \{0.3, 1, 7\}$. As in the case for an instantaneous quench, we observe that thermalization is achieved faster in presence of a non-zero Gauss-Bonnet coupling, but this effect of $\alpha$ on saturation time is attenuated when $t_{q} > \left(1 - 2\alpha\right) z_{\ast}$. We also observe that the maximum rate of growth $\text{max}\left(\mathcal{R}(t) \right)$ increases with decrease in $t_{q}$, and increase in $\alpha$; it may exceed $1$ for very small $t_{q}$. The increase in  $\text{max}\left(\mathcal{R}(t)\right)$ suggests that shorter quenches inject energy into the system more rapidly, resulting in a sharper rise in entanglement entropy. In contrast, for larger $t_{q}$, the system undergoes a smoother transition, leading to a lower peak growth rate. We also note that change in $\frac{\mu}{T}$ does not influence the thermalization as shown in figure \ref{HEE evolution wrto CFT parameter linaer}.

\subsubsection{Periodic quench} \label{sec:3.2.3}

As a final example, we consider a periodically driven system. The goal is to obtain analytic results for the evolution of entanglement entropy using a quench profile which oscillates in time. We consider
\begin{equation} \label{Periodic quench profile}
	g(v)= \sin (\omega  v )\ \Theta(v), 
\end{equation}
where $\omega$ is the driving frequency of the quench. Once again, using equation \eqref{entropy change 0} and \eqref{eq:L1_final} we can calculate the change in entropy for such a quench. In this context, the evolution of entanglement entropy can be categorized into three stages: the pre-quench stage (\(t < 0\)), the initial stage (\(0 < t < \left(1-2\alpha\right) z_{\ast} \)) and the fully driven stage $\left(t > \left(1-2\alpha \right) z_{\ast}\right)$, where
\begin{equation} \label{Entropy of periodic quench}
	\Delta S_A(t) = 
	\begin{cases}
		0, & t < 0, \\ \frac{\ell_{\perp}^2 z_{\ast}^2 \varepsilon}{4 G_{N} z_{H}^4} \int_{0}^{t}dz \ z\sqrt{1-\frac{z^6}{z_{*}^6}} \ \sin \left(\omega \left(t - \left(1 - 2\alpha \right) z \right) \right) \left(1 - 5\alpha \right),
		& 0 < t < \left(1-2\alpha \right) z_{\ast}, \\
		\frac{\ell_{\perp}^2 z_{\ast}^2 \varepsilon}{4 G_{N} z_{H}^4} \int_{0}^{z_{*}}dz \ z\sqrt{1-\frac{z^6}{z_{*}^6}} \ \sin \left(\omega\left(t- \left(1 - 2\alpha \right) z \right) \right)(1-5\alpha), & t > \left(1-2\alpha\right) z_{\ast}.
	\end{cases}
\end{equation}
The initial stage represents a brief transient period. Our primary focus will be on the fully driven phase. Hence we assume \(t > \left(1-2\alpha \right) z_{\ast}\) for the remainder of this section. In this regime, it has been observed that holographic entanglement entropy follows the dynamics of a simple harmonic oscillator \cite{Lokhande:2017jik}. By differentiating equation \eqref{Entropy of periodic quench} twice with respect to \(t\), we arrive at
\begin{equation}
	\frac{d^2(\Delta S_{A})}{dt^2}+\omega^2\Delta S_{A}=0,
\end{equation}
which admits the solution  
\begin{equation}\label{Harmonic sol}
	\Delta S_A(t) = \mathcal{A}(\omega) \sin(\omega t + \phi_A(\omega)).
\end{equation}
It is possible to obtain analytical expressions for \(\mathcal{A}(\omega)\) and \(\phi(\omega)\) for the entangling geometry of our interest. To simplify things, let us first rewrite equation \eqref{Harmonic sol} in the form  
\begin{equation} \label{eq:psi+chi}
	\Delta S_A(t) = \psi_A(\omega)\cos(\omega t) + \chi_A(\omega)\sin(\omega t),
\end{equation} 
and define the amplitude and phase as  
\begin{equation}
	\mathcal{A}(\omega) = \sqrt{\psi_A^2(\omega) + \chi_A^2(\omega)}, \quad \phi(\omega) = \tan^{-1}\left(\frac{\psi_A(\omega)}{\chi_A(\omega)}\right).
\end{equation}
In the case of a strip subsystem, closed-form expressions are obtained in terms of hypergeometric functions
\begin{multline}
	\chi_{\text{strip}} = \frac{\ell_{\perp}^2 z_{*}^4 \varepsilon}{4 G_N z_{H}^4} \frac{1}{216} \left( \frac{18 \sqrt{\pi} \Gamma\left(\frac{1}{3}\right) \, {}_0F_5\left[\frac{1}{6}, \frac{1}{2}, \frac{2}{3}, \frac{5}{6}, \frac{11}{6}, -\left(\frac{(1 - 2 \alpha) \omega z_{*}}{6}\right)^6\right]}{\Gamma\left(\frac{11}{6}\right)}  \right. \\
	- \frac{9 (1 - 2 \alpha)^2 \sqrt{\pi} \omega^2 z_{*}^2  \Gamma\left(\frac{2}{3}\right) \, {}_0F_5\left[\frac{1}{2}, \frac{5}{6}, \frac{7}{6}, \frac{4}{3}, \frac{13}{6}, -\left(\frac{(1 - 2 \alpha) \omega z_{*}}{6}\right)^6\right]}{\Gamma\left(\frac{13}{6}\right)} \\
	+ \left. (1 - 2 \alpha)^4 \omega^4 z_{*}^4 \, {}_1F_6\left[1, \frac{5}{6}, \frac{7}{6}, \frac{4}{3}, \frac{3}{2}, \frac{5}{3}, \frac{5}{2}, -\left(\frac{(1 - 2 \alpha) \omega z_{*}}{6}\right)^6\right] \right)\left(1-5\alpha\right)
\end{multline}
\begin{multline}
	\psi_{\text{strip}} = -\frac{\ell_{\perp}^2 z_{*}^5 \varepsilon}{4 G_N z_{H}^4}\frac{1}{5184} \sqrt{\pi} \omega (1-2\alpha) \Bigg( 432  \sqrt{\pi}  \, {}_0F_5\left[\frac{1}{3}, \frac{2}{3}, \frac{5}{6}, \frac{7}{6}, 2, -\left(\frac{(1 - 2 \alpha)\omega z_{*} }{6}\right)^6\right] \\
	- \frac{12 \sqrt{3} (1 - 2 \alpha)^2 z_{*}^2 \omega^2 \Gamma\left(-\frac{4}{3}\right) \, {}_0F_5\left[\frac{2}{3}, \frac{7}{6}, \frac{4}{3}, \frac{3}{2}, \frac{7}{3}, -\left(\frac{(1 - 2 \alpha) \omega z_{*}}{6}\right)^6\right]}{\Gamma\left(\frac{7}{6}\right)} \\
	+ \frac{\pi (1 - 2 \alpha)^4 \omega^4 z_{*}^4 \, {}_0F_5\left[\frac{4}{3}, \frac{3}{2}, \frac{5}{3}, \frac{11}{6}, \frac{8}{3}, -\left(\frac{(1 - 2 \alpha) \omega z_{*}}{6}\right)^6\right]}{\Gamma\left(\frac{11}{6}\right) \Gamma\left(\frac{8}{3}\right)} \Bigg)\left(1-5\alpha\right)
\end{multline}
\begin{figure}[t]
	\centering
	\begin{subfigure}[b]{0.48\textwidth}
		\centering  
		\includegraphics[width=\textwidth]{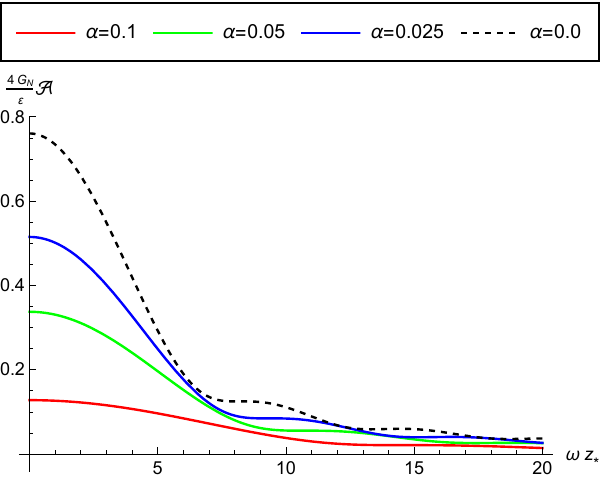}
		\caption{}
		\label{fig:amplitude_periodic}
	\end{subfigure}
	\hfill
	\begin{subfigure}[b]{0.51\textwidth}
		\centering  
		\includegraphics[width=\textwidth]{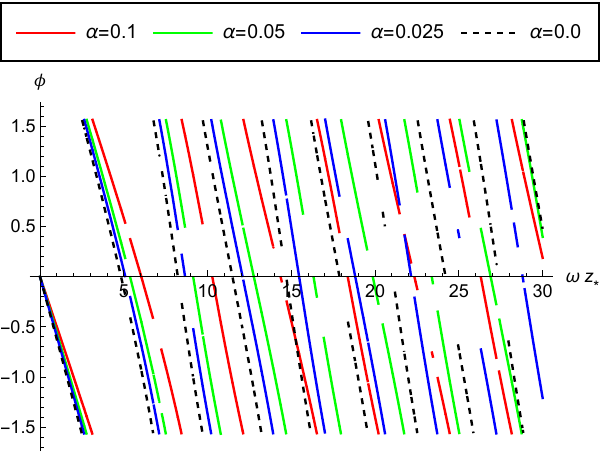}
		\caption{}
		\label{fig:phase_periodic}
	\end{subfigure}
	\caption{Amplitude $\mathcal{A}$ and relative phase $\phi$ as a function of $\omega z_{*}$ for a strip in 5-dimensional Gauss-Bonnet gravity. Both amplitude and phase are presented for different values of the Gauss-Bonnet coupling ($\alpha$). Increasing $\alpha$ decreases the amplitude, while the phase shift remain within $\left(-\frac{\pi}{2}, \frac{\pi}{2}\right)$.}
	\label{fig:Amp and phase periodic}
\end{figure}
A natural next step is to study how \(\mathcal{A}\) and \(\phi\) vary as functions of \(\omega\) during a periodically driven quench in Gauss-Bonnet gravity. In Figure \eqref{fig:Amp and phase periodic}, we present the amplitudes and relative phases for a strip subregion, corresponding to different permissible values of the Gauss-Bonnet coupling ($\alpha$). In all cases, the amplitudes peak as \(\omega \to 0\) and gradually decay as \(\omega \to \infty\), exhibiting mild oscillations at intermediate frequencies. Again, the effect of Gauss-Bonnet coupling ($\alpha$) is such that as $\alpha$ increases, the amplitude decreases. The decrease in amplitude with increasing $\alpha$ suggests that Gauss-Bonnet corrections introduce an effective damping mechanism, reducing the magnitude of entanglement oscillations. In the case of a strip in 5-dimensional Gauss-Bonnet gravity, the phase \(\phi\) varies within the range \(\phi \in (-\pi/2, \pi/2)\).

Let us recall that a periodic source, as given in equation \eqref{Periodic quench profile}, is unphysical since it violates the Null Energy Condition (NEC) \cite{Caceres:2013dma, Callan:2012ip}. In accordance with the laws of black hole thermodynamics, the mass of a black hole must be non-decreasing. To make equation \eqref{Periodic quench profile} non-decreasing, one may introduce a monotonically increasing source. A simple example of such a source is the addition of a linear pump \cite{Lokhande:2017jik}
\begin{equation}
	g(v)= [\sin (\omega  v )+\eta \ t]\ \Theta(v) \ ,\eta\geq\omega
\end{equation}
\begin{figure}[t]
	\centering
	\begin{subfigure}[b]{0.48\textwidth}
		\centering  
		\includegraphics[width=\textwidth]{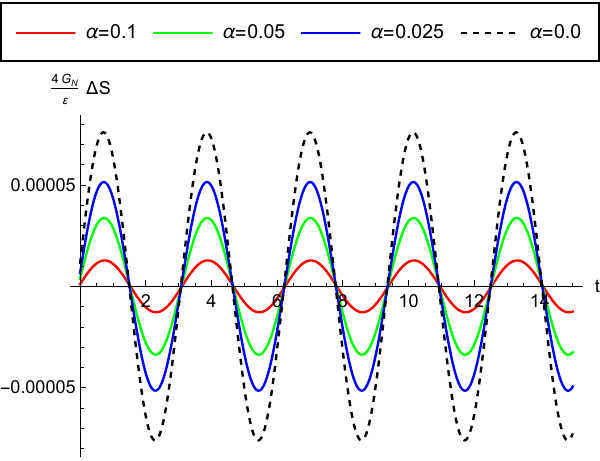}
		\caption{}
		\label{fig:entropy_periodic_asitis}
	\end{subfigure}
	\hfill
	\begin{subfigure}[b]{0.45\textwidth}
		\centering  
		\includegraphics[width=\textwidth]{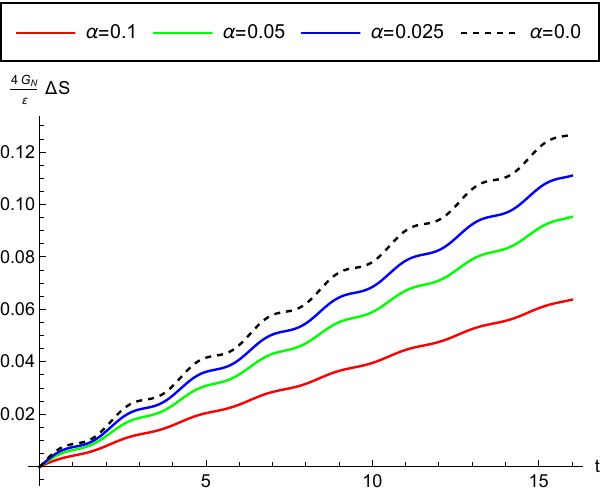}
		\caption{}
		\label{fig:entropy_periodic_withsource}
	\end{subfigure}
	\caption{Evolution of entanglement entropy $\Delta S_A(t)$ for a strip like region in 5-dimensional Gauss-Bonnet gravity comparing strictly periodic driving (left) with a modified NEC-respecting source (right) with $\eta=4$. Curves are obtained for $\omega=2$ and different $\alpha$.}
	\label{fig:Entropy for periodic quench strip}
\end{figure}
The entire entanglement evolution, including both the cases with and without linear driving, is depicted for sample parameters in Figure \eqref{fig:Entropy for periodic quench strip}. It is worth noticing that entanglement entropy monotonically increases for the case where the external quench source respects the NEC as mentioned in \cite{Lokhande:2017jik}.

\subsection{Holographic entanglement entropy of the sphere} \label{subsec 3.3}

We can analyse the entanglement associated with a spherical boundary subregion by expanding the area functional in equation \eqref{eq:functional-ball} following the perturbative approach as in the strip case. Using equations \eqref{F fn quench} and \eqref{eq:v_soln_strip}, we derive the zeroth-order Lagrangian functional
\begin{equation}
	\mathcal{L}^{(0)} = \frac{r^2(z)\sqrt{2 r'(z)^2 + 1 + \sqrt{1-8 \alpha }}}{\sqrt{2} z^3} \left(1 + \mathsf{L}_{\text{sphere}} \left[r'(z),r''(z)\right] \right)
\end{equation}
We avoid writing the expression for \( \mathsf{L}_{\text{sphere}} \) here explicitly since it is somewhat cumbersome. We can derive the equation of motion using the Euler-Lagrange approach. However, unlike the strip case (see equation \eqref{eq:L0_strip_level1}), the Lagrangian for a sphere explicitly depends on \( r(z) \), so we no more have access to a first integral of motion, and the Euler-Lagrange equation has to be solved explicitly. Once again, to maintain analytical control, we restrict our analysis to small $\alpha$, ensuring that higher-derivative corrections remain perturbative.

To avoid clutter, we do not present the specific form of the equation of motion, directly going to the perturbative solution at the leading order. For \( \alpha \ll 1 \), the solution for the Ryu-Takayanagi extremal hypersurface \( r(z) \) takes the following form
\begin{equation}\label{r solution sphere}
	r(z)=\sqrt{1-z^2}(1+3\alpha)+\mathcal{O}(\alpha^2).
\end{equation}
Although this was obtained in units of $z_{*}=1$, we can easily obtain the relation between turning point $z_{*}$ and radius of the sphere R as 
\begin{equation}\label{turning point sphere}
	z_{*}=\frac{L_{\text{eff}}R}{(1+3\alpha)}\simeq R(1-4\alpha).
\end{equation}
Recall that $L_{\text{eff}}$ is the effective radius of the asymptotically AdS spacetime given in \eqref{eq:AdS_Leff}, and the radius of the dual CFT subsystem is $L_{\text{eff}} R$. We use this solution to calculate the leading order change in holographic entanglement entropy by subtracting the zeroth order terms
\begin{equation}
	\label{entropy change 0 sphere}
	\Delta S_{A}(t) = \frac{\Delta \mathcal{A}(t)}{4 G_{N}} = \frac{L_{\text{eff}}^2 A_{2} \lambda^4}{4 G_{N}} \int_{0}^{1} dz \, \left. \mathcal{L}^{(1)}\left[\phi_{A}^{(0)}(z)\right]\right|_{\phi_{A}^{(0)}(z) = r(z)},
\end{equation}
where the on-shell first order change of the Lagrangian up to $\mathcal{O}\left(\alpha\right)$, upon using equations \eqref{eq:v_soln_strip} and \eqref{r solution sphere} can be written as
\begin{equation} \label{eq:L1_final sphere}
	\mathcal{L}^{(1)}\left[r(z)\right] = \frac{z {(1-z^2)^{\frac{3}{2}}}}{2} \left(1 - 3\alpha \right) \varepsilon g \left(t - \frac{z}{\frac{1 - \sqrt{1 - 8\alpha}}{4\alpha}} \right).
\end{equation}
As previous, we study the growth of entanglement entropy and the rate of thermalization for three different quench profiles. Since the entanglement entropy analysis follows a similar structure to the strip case, we summarize the key results without repeating derivations in full detail.

\subsubsection{Instantaneous quench}

We substitute the Lagrangian for a spherical subregion given in \eqref{eq:L1_final sphere} in equation \eqref{entropy change 0 sphere} to obtain
\begin{equation}
	\Delta S_{A}(t)=\frac{L^2_{\text{eff}}A_{2}\lambda^4\varepsilon}{8G_{N}}\int_{0}^{1}z(1-z^2)^{\frac{3}{2}} \left(1-3\alpha \right) \Theta\left(t-\frac{z}{\frac{1-\sqrt{1-8\alpha}}{4\alpha}}\right)
\end{equation}
\begin{equation}
	\simeq\frac{A_{2}\lambda^4\varepsilon}{8G_{N}}\int_{0}^{1}z(1-z^2)^{\frac{3}{2}}(1-5\alpha) \Theta\left(t-(1-2\alpha)z\right)
\end{equation}
which resembles equation \eqref{Entropy change} and can be evaluated in a similar way. The upshot of the calculation is that
\begin{equation} \label{Change in entropy Delta S_A sphere}
	\Delta S_{A}(t) = \Delta S_{\text{eq}}\{\left[\Theta(t) - \Theta\left(t - t_{\text{sat}} \right) \right] \mathcal{G}\left(\frac{t}{t_{\text{sat}}} \right) + \Theta \left(t - t_{\text{sat}} \right)\},
\end{equation}
where
\begin{equation}\label{tsat}
	\Delta S_{\text{eq}} = \frac{\varepsilon A_{2}}{40G_{N}} \frac{z_{*}^2}{z_{H}^4} \left(1-5\alpha \right),\quad t_{\text{sat}}=z_{*}(1-2\alpha),
\end{equation}
and $\mathcal{G}$ is given by
\begin{equation}\label{exp of G(t) sphere}
	\mathcal{G}(x) = \left(1- \left(1-x^2 \right)^{\frac{5}{2}}\right),\quad x = \frac{t}{t_{\text{sat}}}.
\end{equation}
By construction the function \( \mathcal{G}(x) \) takes on the extreme values
\begin{equation}
	\mathcal{G}(0) = 0,\quad \text{and}\quad \mathcal{G}(1) =  1. 
\end{equation}
Hence its average rate of change over this interval is $\left\langle \frac{d\mathcal{G}(x)}{dx} \right\rangle = 1$. We can now determine the instantaneous rate of entanglement growth
\begin{equation} \label{Rate eqn sphere}
	\mathcal{R}(t) \equiv \frac{V_{2}}{A_{2}t_{\text{sat}}} \frac{d\mathcal{G}}{dx} = \frac{5}{3(1-2\alpha)^2(1-4\alpha)}\frac{t}{t_{\text{sat}}}\left(1-\left(\frac{t}{t_{\text{sat}}}\right)^2\right)^{\frac{3}{2}}.
\end{equation}
The time-averaged entanglement velocity is similarly defined as
\begin{equation}
	v_{E}^{\text{avg}} = \langle \mathcal{R}(t) \rangle = \frac{V_{2}}{A_{2} t_{\text{sat}}} \left\langle \frac{d\mathcal{G}(x)}{dx} \right\rangle = \frac{V_{2}}{A_{2} t_{\text{sat}}} =\frac{(1+4\alpha)}{3} + \mathcal{O}\left(\alpha^2\right).
\end{equation}
To obtain the last expression, we used equation \eqref{turning point sphere} and \eqref{tsat} to relate the turning point $z_{*}$ with $t_{\text{sat}}$.
\begin{figure}[t]
	\centering
	\begin{subfigure}[b]{0.48\textwidth}
		\centering  
		\includegraphics[width=\textwidth]{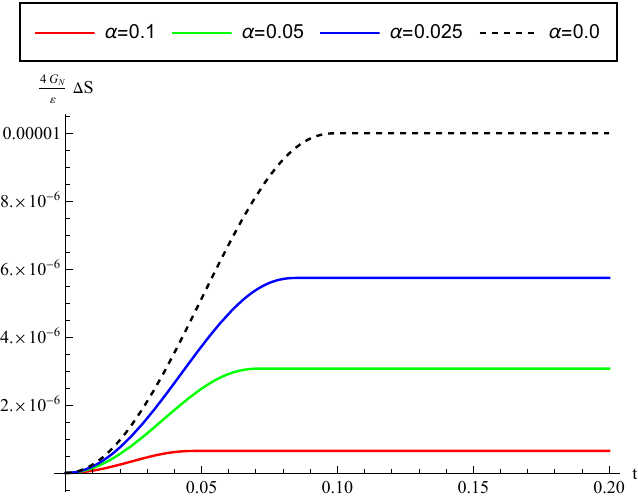}
		\caption{}
		\label{fig:HEE_evolution_instant sphere}
	\end{subfigure}
	\hfill
	\begin{subfigure}[b]{0.48\textwidth}
		\centering  
		\includegraphics[width=\textwidth]{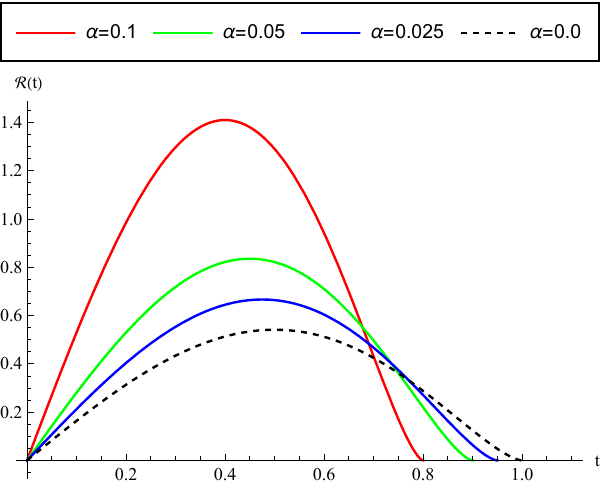}
		\caption{}
		\label{fig:Rate_instant sphere}
	\end{subfigure}
	\caption{Evolution of holographic entanglement entropy for a sphere in five-dimensional Gauss-Bonnet gravity, with $\frac{R}{z_{h}}=0.1$, and  $\frac{A_{2}}{4G_{N}}=1$. (a) Entanglement entropy growth after an instantaneous global quench for different values of $\alpha$, (b) the instantaneous growth-rate $\mathcal{R}(t)$, we observe that the maximum growth-rate in sensitive to $\alpha$ and may exceed $1$.}
	\label{fig:instantaneous_ball}
\end{figure}

Our results are summarised in figure \ref{fig:instantaneous_ball}. As before, we observe that Gauss-Bonnet corrections leads to faster thermalization of the subsystem, accompanied by a higher instantaneous growth rate. When $\alpha$ is decreased, the growth rate also decreases and thermalization is delayed. Consequently, the final state exhibits greater entanglement with its complement. This suggests that the influence of Gauss-Bonnet coupling on thermalization is at least qualitatively universal, independent of the shape of the entangling subregion. 

\subsubsection{Linear quench}
\begin{figure}[t]
	\centering
	\begin{subfigure}{0.32\textwidth}
		\includegraphics[width=\textwidth]{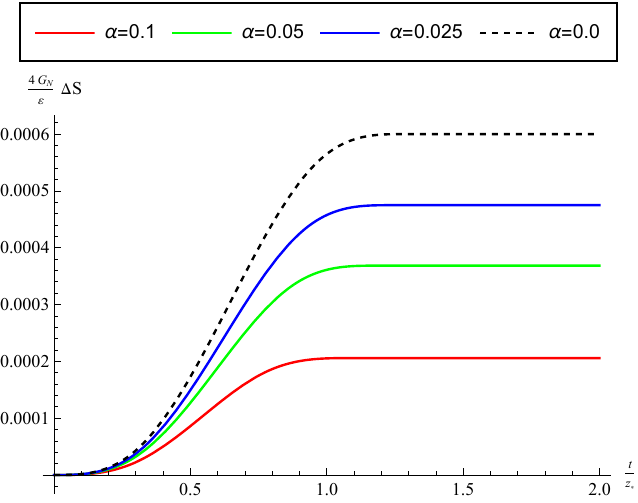}
		\caption{}
	\end{subfigure}
	\hfill
	\begin{subfigure}{0.32\textwidth}
		\includegraphics[width=\textwidth]{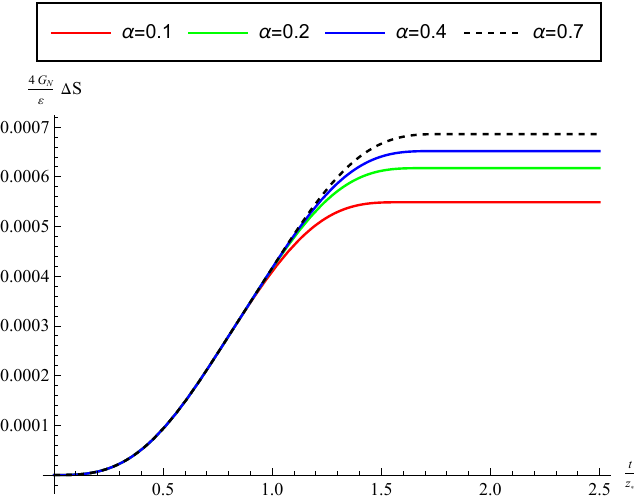}
		\caption{}
	\end{subfigure}
	\begin{subfigure}{0.32\textwidth} 
		\includegraphics[width=\textwidth]{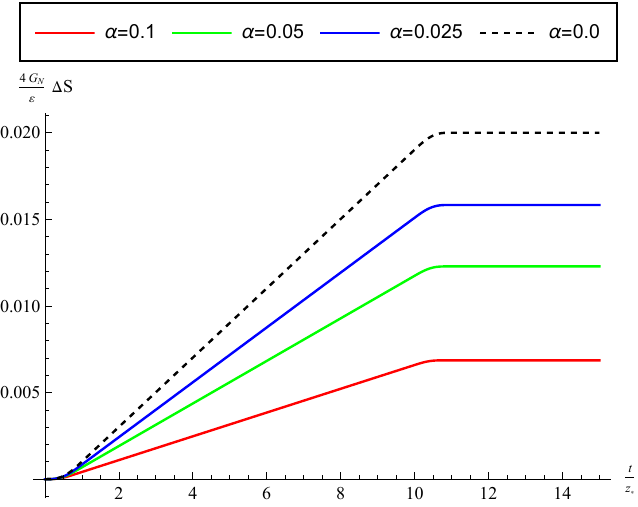}
		\caption{}
	\end{subfigure}
	\hfill
	\begin{subfigure}[b]{0.32\textwidth} 
		\includegraphics[width=\textwidth]{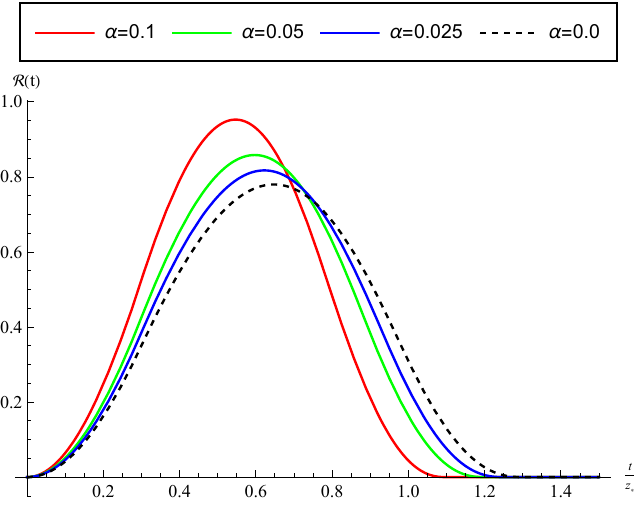}
		\caption{}
	\end{subfigure}
	\begin{subfigure}[b]{0.32\textwidth}
		\includegraphics[width=\textwidth]{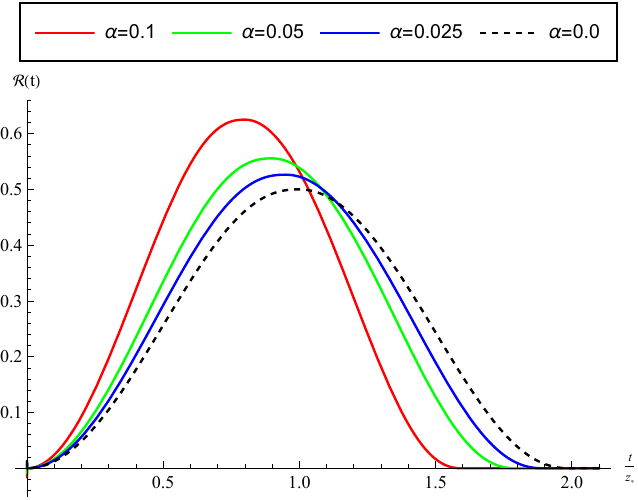}
		\caption{}
	\end{subfigure}
	\hfill
	\begin{subfigure}[b]{0.32\textwidth}  
		\includegraphics[width=\textwidth]{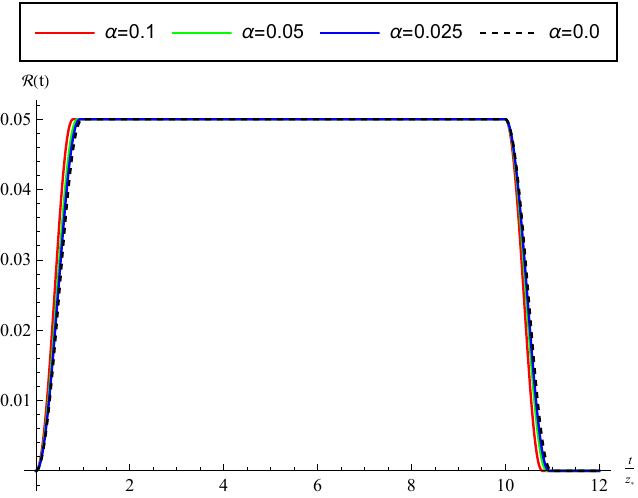}
		\caption{}
	\end{subfigure}
	\caption{Entanglement entropy $\Delta S(t)$ and rate of growth $\mathcal{R}(t)$ for a ball subregion with $\frac{R}{z_{h}} = 0.1$ after a linearly driven quench. Plots are drawn for $t_q={(0.3, 1-2\alpha, 7)}$ from left to right. In all plots we consider $A_{2}$ =1.}
	\label{fig:power law all plot for sphere}
\end{figure}

The linear quench profile is given by equation \eqref{Power-Law quench} for $p=1$. The evolution of entanglement entropy can be broken into the regimes as discussed previously in table \ref{table:quenches}. In analogy with \eqref{eq:strip_indef_int_A} and \eqref{eq:strip_indef_int_B}, to systematically compute the entanglement entropy evolution for a spherical subregion, we define two indefinite integrals
\begin{subequations}
	\begin{align}
		\mathcal{I_{M}}(t,z) &= \frac{A_{2} z_{*}^2\varepsilon}{8G_{N}z_{H}^4} \int dz\ z\left({1-\frac{z^2}{z_{*}^2}}\right)^{\frac{3}{2}}\ \left(t - \left(1 - 2\alpha \right)z \right) \left(1 - 5\alpha \right),\\
		\mathcal{I_{N}}(t_q,z) &= \frac{A_{2} z_{*}^2\varepsilon}{8G_{N}z_{H}^4} \int dz\ z\left({1-\frac{z^2}{z_{*}^2}}\right)^{\frac{3}{2}}\ t_{q} \left(1 - 5\alpha \right).
	\end{align}
\end{subequations} 
Using these two integrals change in entanglement entropy $\Delta S_{A}(t)$ associated with a sphere can be expressed as
\begin{equation}\label{integral for t less for sphere}
	\Delta S_A^{(I)}(t) = 
	\begin{cases}
		0, & t < 0, \\
		\mathcal{I_{M}}(t,z)  |^{\frac{t}{(1-2\alpha)}}_{0}, & 0 < t < t_q, \\
		\mathcal{I_{N}}(t_q,z)  |^{\frac{t-t_q}{(1-2\alpha)}}_{0} +   \mathcal{I_{M}}(t,z)  |^{\frac{t}{(1-2\alpha)}}_{\frac{t-t_q}{(1-2\alpha)}}, & t_q < t < (1-2\alpha)z_{\ast}, \\
		\mathcal{I_{N}}(t_q,z)  |^{\frac{t-t_q}{(1-2\alpha)}}_{0} +   \mathcal{I_{M}}(t,z)  |^{z_{\ast}}_{\frac{t-t_q}{(1-2\alpha)}}, & \left(1-2\alpha\right)z_{\ast} < t < t_{\text{sat}}, \\
		\mathcal{I_{N}}(t_q,z)  |^{z_{\ast}}_{0}, & t > t_{\text{sat}}.
	\end{cases}
\end{equation}
and 
\begin{equation}\label{integral for t greater for sphere}
	\Delta S_A^{(II)}(t) = 
	\begin{cases}
		0, & t < 0, \\
		\left. \mathcal{I_{M}}(t,z) \right|^{\frac{t}{(1-2\alpha)}}_{0}, & 0 < t < (1-2\alpha)z_{\ast}, \\
		\left. \mathcal{I_{M}}(t,z) \right|^{z_{\ast}}_{0}, & (1-2\alpha)z_{\ast} < t < t_q, \\
		\left. \mathcal{I_{N}}(t_q,z) \right|^{\frac{t-t_q}{(1-2\alpha)}}_{0} + \left. \mathcal{I_{M}}(t,z) \right|^{1}_{\frac{t-t_q}{(1-2\alpha)}}, & t_q< t < t_{\text{sat}}, \\
		\left. \mathcal{I_{N}}(t_q,z) \right|^{z_{\ast}}_{0}, & t > t_{\text{sat}}.
	\end{cases}
\end{equation}
The evolution of entanglement entropy $\Delta S_{A}(t)$ and its instantaneous growth rate $\mathcal{R}(t)$ are depicted in figure \ref{fig:power law all plot for sphere} for different quench duration, i. e. different values of $t_{q}$. The behaviour is similar to that observed for a strip.

\subsubsection{Periodic quench}
Finally we turn to periodic quenches for a ball subregion. Following the same method mentioned in section \ref{sec:3.2.3}, we can find out the holographic entanglement entropy in the fully driven phase by solving an equation similar to the harmonic oscillator. For the spherical subsystem we find that
\begin{subequations}
	\begin{align}
		\psi_{\text{sphere}} &= -\frac{A_{2}z^2_{*}\varepsilon}{8G_{N}z_{H}^4}\frac{3\pi\, J_{3} \left( \left(1-2\alpha \right)\omega z_{*} \right)}{2\omega^2(1-2\alpha)^2}, \\
		\chi_{\text{sphere}} &= \frac{A_{2}z^2_{*}\varepsilon}{8G_{N}z_{H}^4}\left(\frac{z_{*}^2}{5}-\frac{3\pi\, H_{3} \left( \left(1-2\alpha \right)\omega z_{*} \right)}{2\omega^2(1-2\alpha)^2} \right),
	\end{align}
\end{subequations}
where $\psi_{\text{sphere}}$ and $\chi_{\text{sphere}}$ were defined in equation \eqref{eq:psi+chi}. Here, $J_{n}(x)$ is the Bessel function of the first kind, and $H_n(x)$ are the Struve function. They can be expressed in terms of the generalized hypergeometric functions ${}_{1}F_{2}$ and ${}_{0}F_{1}$, respectively, using standard identities.

\begin{figure}[t]
	\centering
	\begin{subfigure}[b]{0.48\textwidth}
		\centering  
		\includegraphics[width=\textwidth]{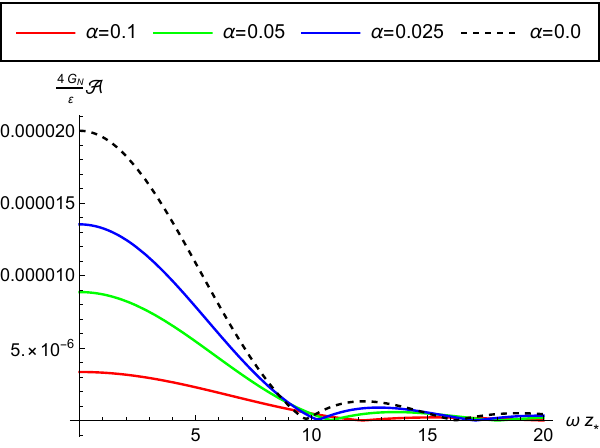}
		\caption{}
		\label{fig:HEE_evolution_instant sphere periodic}
	\end{subfigure}
	\hfill
	\begin{subfigure}[b]{0.48\textwidth}
		\centering  
		\includegraphics[width=\textwidth]{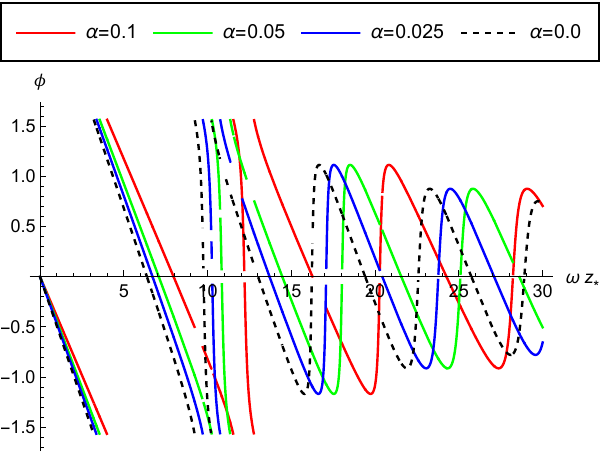}
		\caption{}
		\label{fig:Rate_instant sphere periodic}
	\end{subfigure}
	\begin{subfigure}[b]{0.48\textwidth}
		\centering  
		\includegraphics[width=\textwidth]{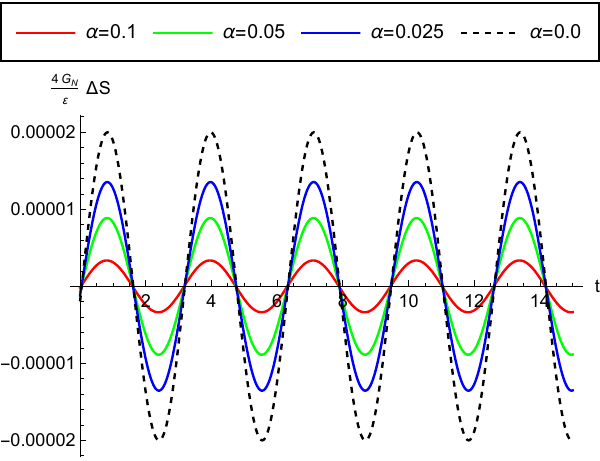}
		\caption{}
		\label{fig:HEE_evolution_instant with linaer source}
	\end{subfigure}
	\hfill
	\begin{subfigure}[b]{0.45\textwidth}
		\centering  
		\includegraphics[width=\textwidth]{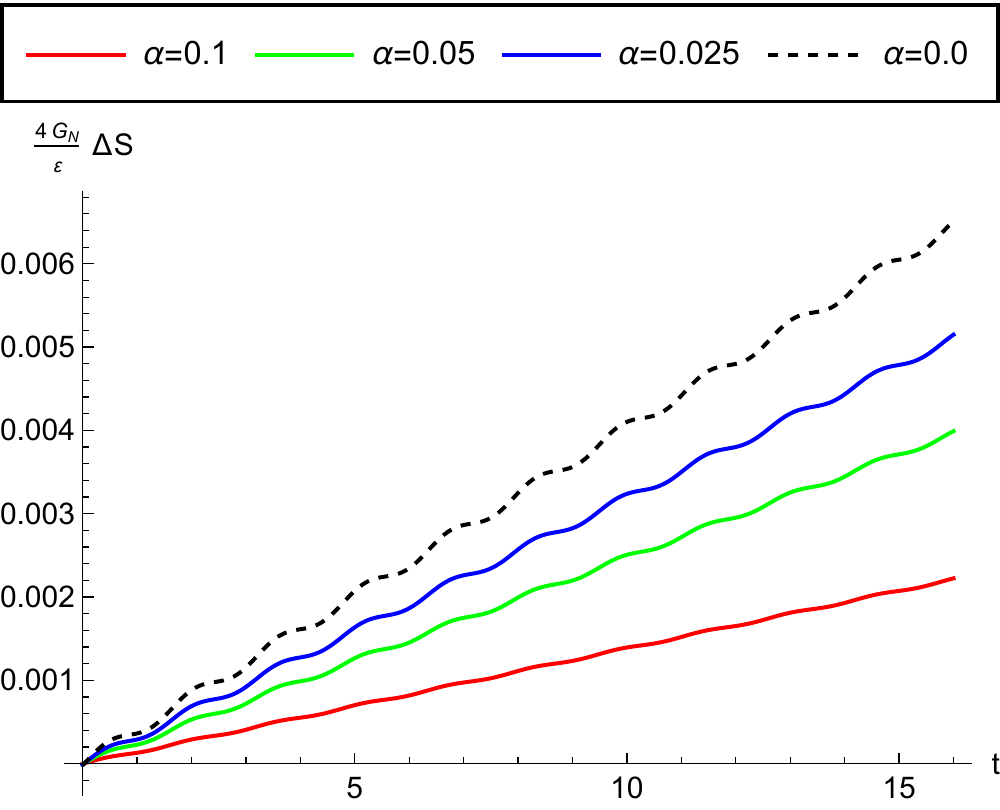}
		\caption{}
		\label{fig:Rate_instant with linear source}
	\end{subfigure}
	\caption{Evolution of entanglement entropy $\Delta S_A(t)$ for a spherical subsystem in 5D Gauss-Bonnet gravity after a periodic quench. Upper panel: (a) amplitude $\mathcal{A} \left(\omega\right)$, and (b) phase $\phi\left(\omega\right)$ of entanglement oscillations as a function of frequency. Lower panel: (c) HEE after a periodic quench, and (d) HEE with an additional linear source with $\eta=4$. In all cases, $\omega = 2$ is fixed.}
	\label{fig:Entropy for periodic quench for sphere}
\end{figure}

We can obtain the behaviour of entanglement entropy as well as the amplitude and phase of $\Delta S_{A}(t)$ easily. They are depicted in figure \ref{fig:Entropy for periodic quench for sphere}.

\section{Regimes of thermalization} \label{sec4}

In this section we expound on the properties of entanglement growth for the two subsystems considered in the previous sections. It has been pointed out in \cite{Kundu:2016cgh, Lokhande:2017jik} that the thermalization process for a small subsystem exhibits three distinct regimes: an initial power-law growth depending on the kind of quench, an intermediate quasi-linear growth, and finally saturation. In the following, we identify these phases for the different kinds of quenches discussed earlier. 

\subsection{Instantaneous quench}

For the instantaneous quench we found that the change in holographic entanglement entropy can be concisely expressed as in equation \eqref{Change in entropy Delta S_A} for the strip and \eqref{Change in entropy Delta S_A sphere} for the sphere. For the convenience of the reader, let us rewrite those results
\begin{equation}\label{Inst all eqn}
		\Delta S_{A}(t) = \Delta S^{\text{eq}}_{A}\{\left[\Theta(t) - \Theta\left(t - t_{\text{sat}} \right) \right] \mathcal{F}_{A}\left(\frac{t}{t_{\text{sat}}} \right) + \Theta \left(t - t_{\text{sat}} \right)\},
\end{equation}
where $\Delta S_{A}^{\text{eq}}$ is the value of entanglement entropy once the system reaches its final equilibrium state
\begin{subequations}
	\begin{align}
		\Delta S^{\text{eq}}_{\text{strip}} &= \frac{\varepsilon \ell_{\perp}^{2}}{4G_{N}} \frac{z_{*}^2}{z_{H}^4} \frac{\sqrt{\pi}}{10} \frac{\Gamma \left(\frac{1}{3} \right)}{\Gamma \left(\frac{5}{6} \right)}\left(1-5\alpha \right),\\
		\Delta S_{\text{ball}}^{\text{eq}} &= \frac{\varepsilon A_{2}}{40G_{N}} \frac{z_{*}^2}{z_{H}^4} \left(1-5\alpha \right),
	\end{align}
\end{subequations}
and $\mathcal{F}_{A}(x)$ is a function that characterizes the growth for different subsystems
\begin{subequations}
	\begin{align}
		\mathcal{F}_{\text{strip}}(x) &= \frac{5\,\Gamma \left(\frac{5}{6} \right) }{\sqrt{\pi}\ \Gamma \left(\frac{1}{3} \right)}\ x^2\, {_2F_1} \left(-\frac{1}{2}, \frac{1}{3}, \frac{4}{3}, x^6 \right),\\
		\mathcal{F}_{\text{ball}}(x) &= \left(1- \left(1-x^2 \right)^{\frac{5}{2}}\right).
	\end{align}
\end{subequations}
In both cases, saturation is attained after $t_{\text{sat}} = \left(1 - 2\alpha \right)z_{*}$ amount of time has passed. Let us now consider the different time regimes during this growth process.		

\noindent \textbf{Initial quadratic growth}

 In the initial phase when $t \ll t_{\text{sat}}$, the growth of entanglement is dominated by the behaviour of \(\mathcal{F}_{A}(x) \) or \( x \ll 1 \). In this limit we find
\begin{subequations}
	\begin{align}
	\mathcal{F}_{\text{strip}}(x) &= \frac{5\Gamma \left(\frac{5}{6} \right)}{\sqrt{\pi}\Gamma \left(\frac{1}{3} \right)} x^2 + \mathcal{O}(x^8),\\
	\text{and}\quad \mathcal{F}_{\text{ball}}(x) &= \frac{5}{2}x^2 + \mathcal{O}(x^4).
	\end{align}
\end{subequations}
Then the early-time growth of the entanglement entropy for both the strip and the sphere is adequately described by
\begin{subequations}
\begin{align}
	\Delta S_{\text{strip}}(t) &\simeq \frac{\varepsilon \ell^2_{\perp}}{8G_{N}} \frac{t^2}{z_{H}^4} \left(1 - \alpha \right), \label{eq:initial_growth_strip} \\
	\Delta S_{\text{ball}}(t) &\simeq \frac{\varepsilon A_{2}}{16G_{N}} \frac{t^2}{z_{H}^4} \left(1 - \alpha \right). \label{eq:initial_growth_ball}
\end{align}
\end{subequations}
Therefore, we conclude that in the early growth regime, when \( t << t_{\text{sat}} = \left(1-2\alpha\right)z_{*} \), the evolution of entanglement entropy is quadratic in time and obeys the proposed universal scaling behaviour \cite{Kundu:2016cgh, Lokhande:2017jik}. Moreover, the absence of any additional geometric quantities, such as \( \ell \) or $R$ in the expressions \eqref{eq:initial_growth_strip} or \eqref{eq:initial_growth_ball} suggests that for small Gauss-Bonnet coupling ($\alpha$) the quadratic growth behaviour \( \Delta S_A(t) \sim t^2 \) might be solely determined by conformal symmetry, as it is for holographic theories dual to Einstein-Hilbert gravity.

We may as well express this universal result in terms of the field theory data $T$ and $\mu$ using equation \eqref{eq:geometry_intermsof_CFT}. In the two extreme limits, we obtain for the strip
\begin{align}
	\begin{split}
		\frac{\mu}{T} \ll 1:\quad \Delta S_{\text{strip}}(t) &\simeq \frac{\varepsilon \ell_{\perp}^2}{8 G_{N}} \pi^4 t^2 \left(1 - \alpha \right) T^4 \left(1 + 128 \frac{\mu^2}{T^2} \right),\\
		\frac{\mu}{T} \gg 1:\quad \Delta S_{\text{strip}}(t) &\simeq \frac{128 \varepsilon \ell_{\perp}^2}{G_{N}} \pi^4 t^2 \left(1 - \alpha \right) \mu^4 \left(1 + \frac{1}{2\sqrt{2}} \frac{T}{\mu} \right).
	\end{split}
\end{align}
While for the sphere
\begin{align}
	\begin{split}
		\frac{\mu}{T} \ll 1:\quad \Delta S_{\text{ball}}(t) &\simeq \frac{\varepsilon A_{2}}{16 G_{N}} \pi^4 t^2 \left(1 - \alpha \right) T^4 \left(1 + 128 \frac{\mu^2}{T^2} \right),\\
		\frac{\mu}{T} \gg 1:\quad \Delta S_{\text{ball}}(t) &\simeq \frac{64 \varepsilon A_{2}}{G_{N}} \pi^4 t^2 \left(1 - \alpha \right) \mu^4 \left(1 + \frac{1}{2\sqrt{2}} \frac{T}{\mu} \right) .
	\end{split}
\end{align}

\noindent \textbf{Quasi-linear growth}

Let $t_{\text{max}}$ denote the instant of time when the growth rate $\mathcal{R}(t)$ reaches its maximum. For intermediate times $t \sim t_{\text{max}}$, for some $0 < t_{\text{max}} < t_{\text{sat}}$, it is possible to define a regime where
\begin{equation} \label{eq:HEE_intermediate_instantaneous}
	\Delta S_{A}(t) - \Delta S_{A}\left(t_{\text{max}}\right) = 2 v_{A}^{\text{max}} s_{\text{eq}} \ell_{\perp}^2 \left(t - t_{\text{max}} \right) + \mathcal{O}\left(t - t_{\text{max}}\right)^3,
\end{equation}
where $v_{A}^{\text{max}}$ is an analogue of the tsunami velocity for large subsystems \cite{Liu:2013iza, Liu:2013qca}, defined to be the maximum rate of growth of entanglement entropy
\begin{equation}
	v_{A}^{\text{max}} \equiv \mathrm{max} \left[\mathcal{R}(t) \right] = \frac{1}{2 s_{\text{eq}}\, \ell_{\perp}^2} \left.\frac{d(\Delta S_{A})}{dt}\right|_{t = t_{\text{max}}}.
\end{equation}
We are discussing here the results for a strip subsystem. The quadratic corrections to equation \eqref{eq:HEE_intermediate_instantaneous} vanish by construction because the rate of growth attains its maxima at $t = t_{\text{max}}$. In the works on holographic entanglement growth in Einstein-Hilbert gravity \cite{Kundu:2016cgh, Lokhande:2017jik}, it was pointed out that contrary to large subsystems \cite{Liu:2013iza, Liu:2013qca}, the equation \eqref{eq:HEE_intermediate_instantaneous} is not a universal relationship. The `velocity' $v_{A}^{\text{max}}$ is not a physical velocity, it is independent of the parameters such as $T$ and $\mu$ which characterize the CFT state, and depends on the shape of the entangling region. It also violates the causality bound $v_{A}^{\text{max}} \leq 1$. Using equation \eqref{Inst all eqn} we find that 
\begin{equation}
	v^{\text{max}}_{E}\rvert_{\text{strip}} \simeq \frac{5\sqrt{3} \Gamma\left(\frac{5}{6} \right) \Gamma\left(\frac{2}{3} \right)}{2^{\frac{1}{3}}\Gamma\left(\frac{1}{3}\right)\Gamma\left(\frac{1}{6}\right)} \left(1+6\alpha \right).
\end{equation}
We can follow similar steps for the case of a sphere. At the end of the computation, we find that in thus case
\begin{equation}
	v^{max}_{E}\rvert_{\text{ball}} \simeq \frac{15\sqrt{3}}{16} \left(1+6\alpha \right).
\end{equation}
giving a lower maximum rate in comparison to the strip. Since $\alpha$ is a positive number, the r.h.s. of both the equations may exceed $1$; this is also evident from figure \ref{fig:Rate_instant} and \ref{fig:Rate_instant sphere}. The only difference between $\alpha = 0$ and $\alpha \neq 0$ cases is that in Einstein-Hilbert gravity $v_{A}^{\text{max}} > 1$ only in $d = 2$, but in presence of a non-zero $\alpha$, the bound is violated even in higher dimensions. Anyway, this is no cause of concern because we have already argued that the physical velocity is the average rate of entanglement growth, which always obeys the causality bound.

\noindent \textbf{Approach to saturation}

By expanding $\mathcal{F}_{A}\left(\frac{t}{t_{\text{sat}}}\right)$ in equation \eqref{Inst all eqn} near $t = t_{\text{sat}}$, we observe that the equilibration of entanglement entropy for the strip subregion resembles a continuous phase transition with
\begin{equation}
	\Delta S_{A}(t) - \Delta S_{\mathrm{eq}} \propto \left(t_{\text{sat}} - t\right)^{\frac{3}{2}}.
\end{equation}
Similarly, by expanding equation \eqref{exp of G(t) sphere} around $t=t_{\text{sat}}$ we also get a non-trivial scaling exponent for the sphere as
\begin{equation}
	\Delta S_{A}(t) - \Delta S_{\mathrm{eq}} \propto \left(t_{\text{sat}} - t\right)^{\frac{5}{2}}.
\end{equation}
This is the same as holographic CFTs dual to Einstein-Hilbert gravity. Thus, the introduction of a Gauss-Bonnet correction does not change the approach to saturation. It does influence the saturation time as we have observed already.
\subsection{Linear quench}
\textbf{Early time growth} 

For the linear quench profile, the early-time growth of entanglement entropy for the strip and ball regions are well approximated by
\begin{subequations}
\begin{align}
	\Delta S_{\text{strip}}(t) \simeq \frac{\ell^2_{\perp}z_{*}^2\varepsilon}{48 G_{N}z_{H}^4} t^3 \left(1 + \alpha \right), \label{eq:initial_growth_linear_strip} \\
	\Delta S_{\text{ball}}(t) \simeq \frac{A_{2}z_{*}^2\varepsilon}{96 G_{N}z_{H}^4} t^3 \left(1 + \alpha \right). \label{eq:initial_growth_linear_ball}
\end{align}
\end{subequations}
For \(p = 1\) and \(\alpha = 0\), this result matches exactly with the findings reported in \cite{Lokhande:2017jik}. Additionally, it reproduces the results for instantaneous quenches discussed in the previous section when we take $p$ arbitrary and consider the limit $p \to 0$. The proof provided in \cite{Kundu:2016cgh} for the universality of early-time growth was based entirely on symmetry arguments and can be readily extended to quenches of finite duration. Using the same reasoning, it is reasonable to conclude that the scaling with time should always hold as long as $0< t \ll t_{\text{sat}}$, regardless of the shape and size of the entangling region.

\noindent \textbf{Intermediate growth} 

From figure \ref{fig:power law all plot} we observe  that the maximum growth rate \(\mathcal{R}(t)\) increases with the Gauss-Bonnet coupling \(\alpha\). The proof that \(v_A^{\text{avg}} \leq 1\), as presented in \cite{Kundu:2016cgh}, holds for a quench of finite duration. As mentioned in \cite{Lokhande:2017jik}, one can think of a finite-duration background \(t_q\) as a collection of thin shells spread across the range \(v \in [0, t_q]\), with the derivation following a similar approach. Similarly, as shown in \cite{Lokhande:2017jik}, we can define a modified average velocity
\begin{equation}
	v_E^{\text{avg}}(t_q, p) = \frac{v_E^{\text{avg}}(t_q \to 0)}{1 + t_q / (1 - 2\alpha)}
\end{equation}
Although we consider \(p = 1\) in our case, the expression is independent of \(p\). It is also clear that, for a constant \(\alpha\), as \(t_q\) increases, the average speed of entanglement propagation decreases, which is also evident from figure \ref{fig:power law all plot}.

\noindent \textbf{Approach to saturation}

Near saturation $t\to t_{\text{sat}}$, the entanglement entropy  for  both strip and sphere is always continuous and resembles a second order phase transition
\begin{equation}
	\Delta S_{A}(t)-\Delta S^{\text{eq}}_{A}\propto (t_{\text{sat}}-t)^{\gamma}.
\end{equation}
where
\begin{equation}
	\gamma_{\text{strip}}=\frac{5}{2},\quad \gamma_{\text{ball}}=\frac{7}{2}
\end{equation}
The exponent can be directly derived from the two integrals in the stage preceding saturation, given by\eqref{integral for t less} and\eqref{integral for t greater} respectively. Similarly, for the case of a sphere this exponent can be calculated from the two integrals \eqref{integral for t greater for sphere} and \eqref{integral for t less for sphere} respectively. As noted in \cite{Lokhande:2017jik}, the above result does not extend to the instantaneous quench.

\section{Linear response of holographic entanglement entropy} \label{sec5}

It is well known that the infinitesimal change in entanglement entropy over the ground state for any time-independent perturbation $\rho_{A} \to \rho_{A} + \Delta\rho_{A}$, obeys a `first law of entanglement' \cite{Wong:2013gua}
\begin{equation*}
	\Delta S_{A} = \text{Tr}_{A}\left(\Delta\rho_{A} H_{A} \right),
\end{equation*}
where $H_{A}$ is the `modular Hamiltonian'. For a spherical region $A$ in a conformal field theory, the modular Hamiltonian can be written explicitly in a local form
\begin{equation*}
	H_{A} = \int_{A} \beta(x) T_{00}(x),
\end{equation*}
where $T_{00}(x)$ is the physical energy density, and $\beta(x)$ is a quantity dubbed the local (inverse) `entanglement' temperature, which depends on the size of the subsystem. When the energy density is uniform over the sphere $A$, the first law of entanglement can be expressed as a thermodynamic relationship
\begin{equation}
	\Delta S_{A} = \beta_{A} \Delta E_{A},\quad \text{with }\ \Delta E_{A} = \langle \Delta T_{00} \rangle,\ \beta_{A} = \frac{\int_{A} \beta(x)}{\text{vol}(A)}.
\end{equation}
This version of the first law is supported by holographic analyses \cite{Bhattacharya:2012mi, Allahbakhshi:2013rda}, which also extend it for a strip subregion.

It is a natural question to inquire if a first law like relationship can be formulated for non-equilibrium states of a holographic CFT. The authors of \cite{Lokhande:2017jik} reasoned that at least for slowly-varying quenches, the relationship
\begin{equation}
	T_{A} \Delta S_{A}(t) = \Delta E_{A}(t),
\end{equation}
is always satisfied as long as $\frac{d\varepsilon(t)}{dt}\ell^{d+1} \ll \varepsilon(t)\ell^{d} \ll 1$. To check how far a system is from satisfying a first law, they introduced the concept of a time-dependent relative entropy
\begin{equation} \label{eq:rel_ent_noneq}
	\Upsilon_{A}(t)=\frac{\Delta E_{A}(t)}{T_{A}}-\Delta S_{A}(t).
\end{equation}
Using very general arguments, it can be shown that \cite{Lokhande:2017jik}
\begin{enumerate}[(i)]
	\item The quantity $\Upsilon_{A}(t)$ becomes zero in equilibrium states, and it is negligible for slowly varying quenches. In the case of a quench with compact support, this implies that \(\Upsilon_A(t) = 0\) for times \(t < 0\) and \(t > t_{\text{sat}}\), where \(t_{\text{sat}}\) represents the saturation time. 
	
	\item The quantity is always non-negative, meaning it must increase initially and then decrease within the interval \(0 < t < t_{\text{sat}}\).
	
	\item For a general quench, regardless of whether it is slow or rapid, this quantity acts as an indicator of how far the out-of-equilibrium state at time \(t\) deviates from an equilibrium state with the same energy density \(\varepsilon(t)\). This behavior is a direct consequence of its definition and the fact that \(\Upsilon_A(t) \geq 0\). 
\end{enumerate}

In this section, we examine the behaviour of $\Upsilon_{A}(t)$ within the setup of our holographic Gauss-Bonnet gravity. For simplicity, we put $Q = 0$, and consider only uncharged black branes. Let us first note that the change in energy and the entanglement temperature in Gauss-Bonnet gravity for a strip subregion can be approximately written as \cite{Guo:2013aca,Bhattacharya:2012mi}
	\begin{subequations}
		\begin{align}
		\Delta E_{A} &= \frac{(d-1) g(v) L_{\text{eff}}^{d-1} \ell^{d-2}}{16\pi G_{N}} \frac{\ell_{\perp}}{z_{h}^d} \left(1 - 4\alpha \right),\\
		T_{A} &= \frac{2 \left(d^{2} - 1 \right) \Gamma\left(\frac{d+1}{2d-2} \right) \Gamma\left(\frac{d}{2d-2} \right)^{2}}{\sqrt{\pi}\,\Gamma\left(\frac{1}{d-1} \right) \Gamma\left(\frac{1}{2d-2}\right)^{2}\, L_{\text{eff}}\ell_{\perp}} \left(1 - 3\alpha\right).
		\end{align}
	\end{subequations}
For a spherical subregion the relations are 
	\begin{subequations}
		\begin{align}
			\Delta E_{A} &= \frac{g(v) A_{2}  }{16\pi G_{N}z^d_{h}}(1-4\alpha),\\
			T_{A} &= \frac{d+1}{2\pi L_{\text{eff}}R}.(1-3\alpha)
		\end{align}
	\end{subequations}

\subsection*{Instantaneous quench}

From its definition, it follows that for instantaneous quenches 

\begin{equation}
	\Upsilon_A(t)	= \frac{\Delta E_{A}}{T_A} - \Delta S_A^\text{eq} \mathcal{F}_A\left(\frac{t}{t_\text{sat}}\right), \quad 0 < t < t_\text{sat}
\end{equation}
and \(\Upsilon_A(t) = 0\) otherwise. 

\begin{figure}[t]
	\centering
	\begin{subfigure}[b]{0.48\textwidth}
		\centering  
		\includegraphics[width=\textwidth]{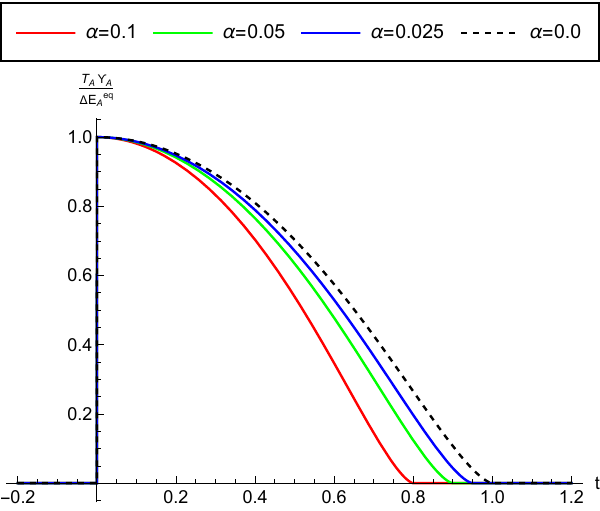} 
		\caption{}
		\label{fig:HEE_evolution_instant_sphere}
	\end{subfigure}
	\begin{subfigure}[b]{0.48\textwidth} 
		\centering  
		\includegraphics[width=\textwidth]{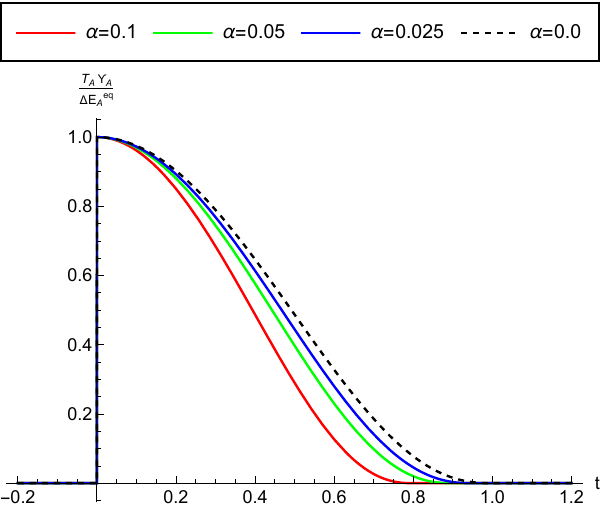} 
		\caption{}
		\label{fig:Rate_instant_sphere}
	\end{subfigure}
	\caption{Time-dependent relative entropy $\Upsilon_{A}(t)$ for (a) the strip subsystem, and (b) the spherical subsystem for different values of Gauss-Bonnet coupling $\alpha$.}
	\label{fig:Relative_entropy_instant}
\end{figure}

Figure \eqref{fig:Relative_entropy_instant} shows the evolution of \(\Upsilon_A(t)\) for instantaneous quenches. These figures align with the expected behaviour discussed earlier. It is noteworthy that, because the quench is instantaneous, the `driven regime' is restricted to the single point \(t = 0\), where \(\Upsilon_A(t)\) increases discontinuously. Finally, it is worth emphasizing that the behaviour of \(\Upsilon_A(t)\) throughout its evolution highlights its role as a measure of the `distance' between the out-of-equilibrium state and the equilibrium state at the same energy density. Specifically, \(\Upsilon_A(t)\) is maximal immediately after the quench and relaxes to zero as \(t \to t_\text{sat}\). A positive Gauss-Bonnet coupling $\alpha$ causes $\Upsilon_{A}$ to decrease -- mimicking its effect on the holographic entanglement entropy.

\subsection*{Linear quench}

\begin{figure}[t]
	\centering
	\begin{subfigure}{0.32\textwidth}
		\includegraphics[width=\textwidth]{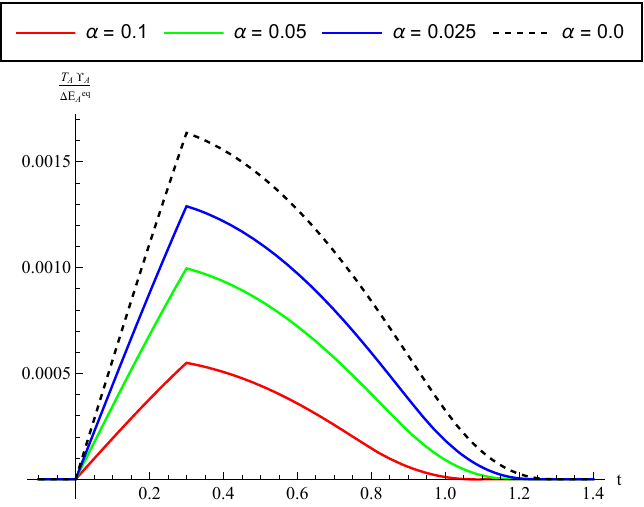}
		\caption{}
	\end{subfigure}
	\hfill
	\begin{subfigure}{0.32\textwidth}
		\includegraphics[width=\textwidth]{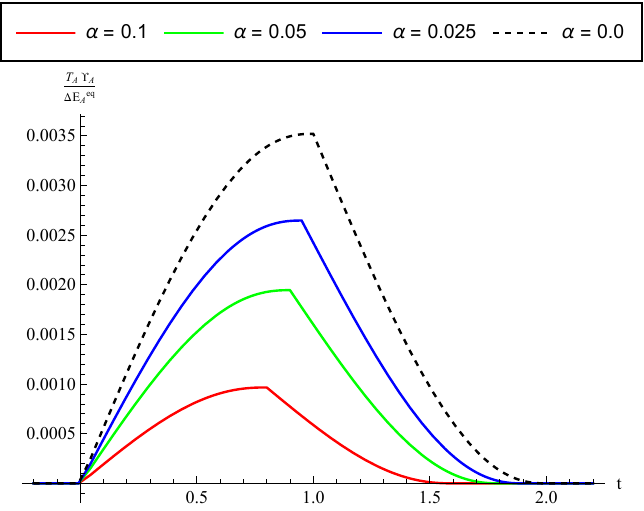}
		\caption{}
	\end{subfigure}
	\begin{subfigure}{0.33\textwidth} 
		\includegraphics[width=\textwidth]{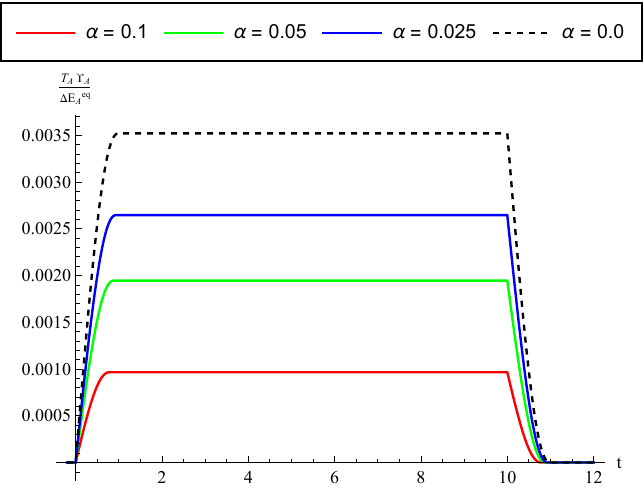}
		\caption{}
	\end{subfigure}
	\caption{$\Upsilon_{A}(t)$ for the strip after a linearly driven quench. We have chosen values of $t_{q}=(0.3,1-2\alpha,10)$ in ascending order from left to right.}
	\label{fig:Relative_entropy_linear_strip}
\end{figure}

\begin{figure}[t]
	\centering
	\begin{subfigure}{0.32\textwidth}
		\includegraphics[width=\textwidth]{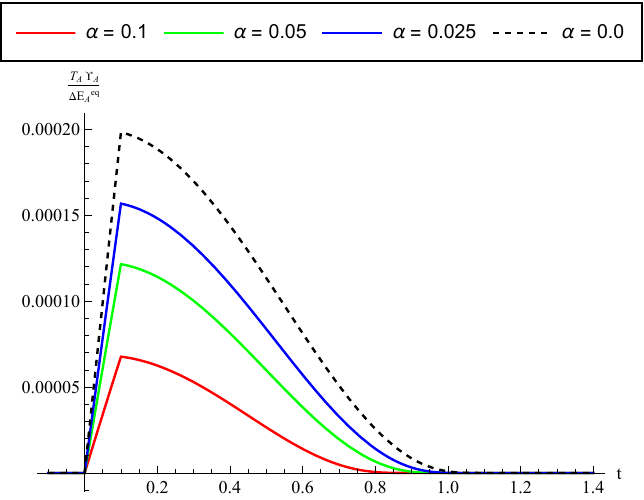}
		\caption{}
	\end{subfigure}
	\hfill
	\begin{subfigure}{0.32\textwidth}
		\includegraphics[width=\textwidth]{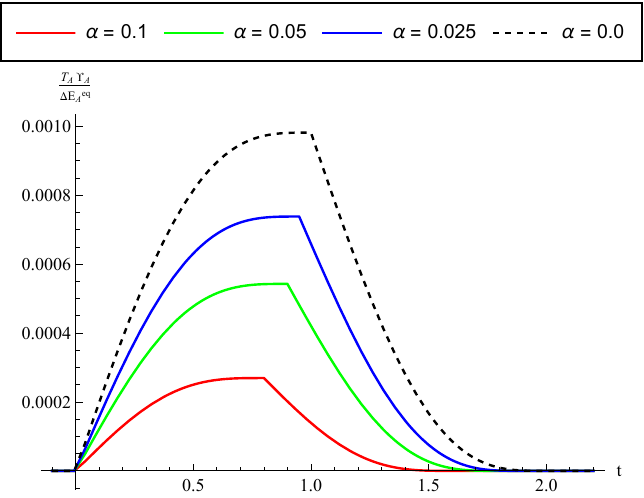}
		\caption{}
	\end{subfigure}
	\begin{subfigure}{0.33\textwidth} 
		\includegraphics[width=\textwidth]{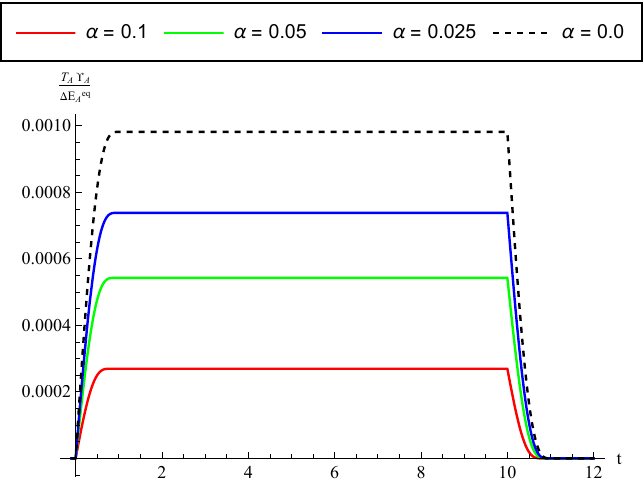}
		\caption{}
	\end{subfigure}
	\caption{Relative entropy $\Upsilon_{A}(t)$ for the sphere after a linearly driven quench with $p=1$. We have chosen $t_{q}/z_{*} = \left(0.3, 1-2\alpha, 10 \right)$ in ascending order from left to right.}
	\label{fig:Relative_entropy_linear_sphere}
\end{figure}

For a quench of finite duration, $\Upsilon_{A}(t)$ increases as long as energy is being injected into the system, reaches its maximum at $t = t_{q}$ when the quench finally ends, and then smoothly approaches zero. The figures \ref{fig:Relative_entropy_linear_strip} and \ref{fig:Relative_entropy_linear_sphere} demonstrate these features for the strip and the sphere, respectively. When $t_{q} \gg \left(1 - 2\alpha \right) z_{*}$, the relative entropy remains constant over a large time. During this period the entanglement evolution is governed by the first law of entanglement rates (FLOER) \cite{OBannon:2016exv}. As usual, a positive $\alpha$ results in a decrease of the overall $\Upsilon_{A}(t)$.
	
\subsection*{Periodic quench}

\begin{figure}[t]
	\centering
	\begin{subfigure}[b]{0.48\textwidth}
		\centering  
		\includegraphics[width=\textwidth]{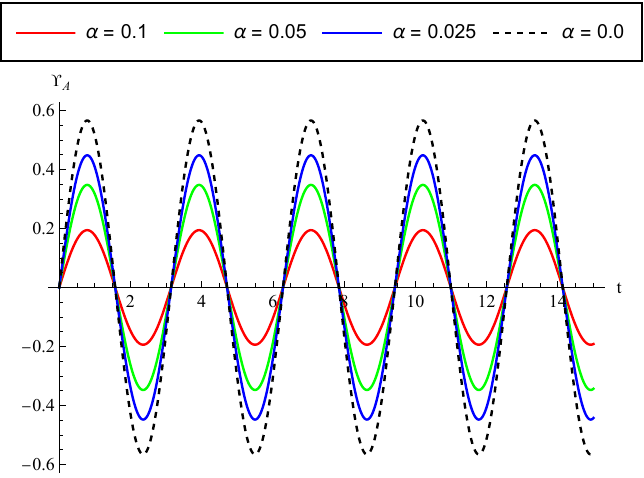}
		\caption{}
		\label{fig:relent_periodic}
	\end{subfigure}
	\hfill
	\begin{subfigure}[b]{0.48\textwidth}
		\centering  
		\includegraphics[width=\textwidth]{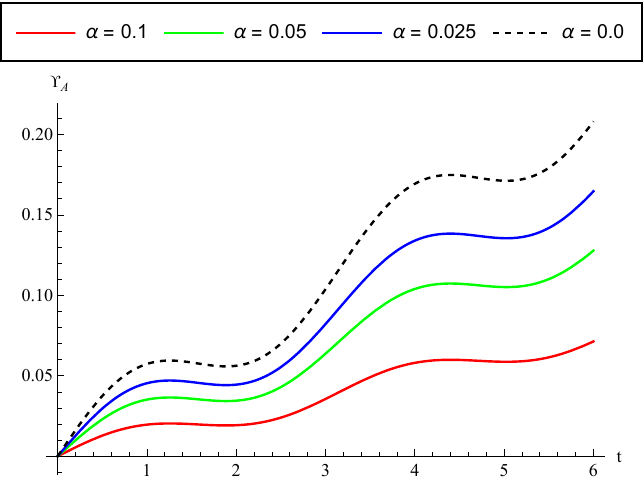}
		\caption{}
		\label{fig:relent_periodic2}
	\end{subfigure}
	\caption{Relative entropy $\Upsilon_{A}(t)$ for a strip like region in 5-dimensional Gauss-Bonnet gravity: (a) strictly periodic source, and (b) source respecting the bulk NEC with $\eta=4$. Curves are obtained with $\omega=2$ and various strength of Gauss-Bonnet coupling ($\alpha$).}
	\label{fig:Entropy for periodic quench}
\end{figure}

For the final example of a periodic quench, we find that the time-dependent relative entropy $\Upsilon_{A}(t)$ usually oscillates around zero for a purely periodic source. In this case $\Upsilon_{A}$ can become negative, but this is expected since the driving does not respect the bulk NEC. After including a linear term - positivity is restored -- as depicted in figure \ref{fig:Entropy for periodic quench}.

\begin{figure}[t]
	\centering
	\begin{subfigure}[b]{0.48\textwidth}
		\centering  
		\includegraphics[width=\textwidth]{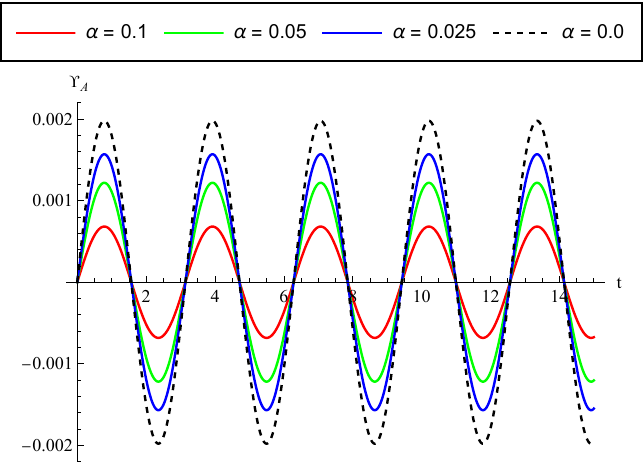}
		\caption{}
		\label{fig:relent_periodic_sphere}
	\end{subfigure}
	\hfill
	\begin{subfigure}[b]{0.46\textwidth}
		\centering  
		\includegraphics[width=\textwidth]{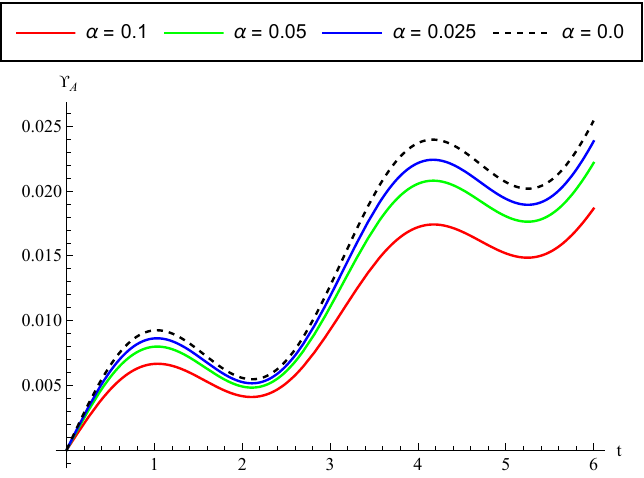}
		\caption{}
		\label{fig:relent_periodic_sphere2}
	\end{subfigure}
	\caption{Relative entropy $\Upsilon_{A}(t)$ for a spherical subsystem in 5-dimensional Gauss-Bonnet gravity is shown. Curves are obtained with $\omega=2$ and various strength of Gauss-Bonnet coupling ($\alpha$). Panel (a) corresponds to strictly periodic source, whereas panel (b) represents a source respecting the bulk NEC with $\eta=4$}
	\label{fig:Entropy for periodic quench sphere}
\end{figure}

\section{Evolution of mutual information} \label{sec6}

As briefly mentioned in the introduction, entanglement entropy exhibits a short-distance divergence due to large amount of entanglement between the degrees of freedom localized near the boundary of the entangling surface. As a result, any computation of entanglement entropy necessitates a regularization scheme. One way to extract a finite, regularization-independent quantity is to construct an appropriate linear combination of entanglement entropies. For a bipartite system, a widely studied example of such a quantity is the mutual information, defined as in equation \eqref{eq:MI_defn}
\begin{equation}
	I\left(A_{1} \cup A_{2} \right) = S_{A_{1}} + S_{A_{2}} - S_{A_{1} \cup A_{2}},
\end{equation}

In this section, we investigate the time evolution of holographic mutual information (HMI) following a global quench. For simplicity, we choose the two regions $A_{1}$ and $A_{2}$ to be infinitely long strips of the same width $\ell$, and separated by a length $h$. Assuming $\frac{\ell}{z_{h}}$, $\frac{h}{z_{h}} \ll 1$, the calculation of mutual information is a straightforward application of the already derived expressions for entanglement entropy in the previous sections, and equation \eqref{eq:MI_defn}.

\begin{figure}[t]
	\centering
	\includegraphics[width=0.75\textwidth]{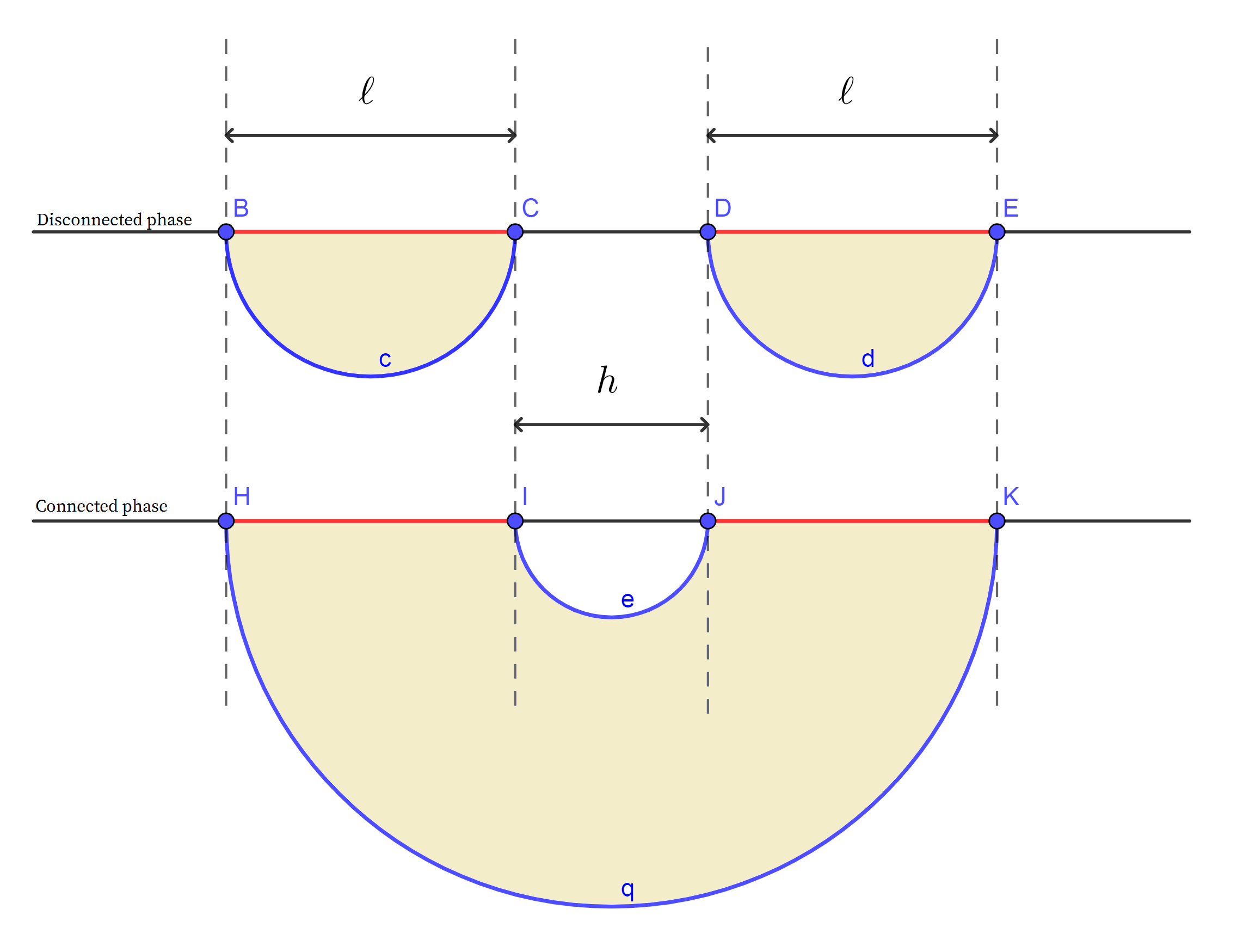}
	\caption{Possible extremal hypersurface for the case of disjoint subregions}
	\label{fig:disentangling}
\end{figure}

It is well known that mutual information exhibits a geometric phase transition as the distance between the two subregions is increased \cite{Headrick:2010zt}. This occurs because in such scenario, there are usually more than one candidate extremal Ryu-Takayanagi hypersurfaces for computing the entanglement entropy $S_{A_{1} \cup A_{2}}$ associated with the union of the two regions. The entropy is determined by the hypersurface with the minimum area. In the static case, the two competing candidates are shown in figure \ref{fig:disentangling}: (1) A `connected' hypersurface bridging $A_{1}$ and $A_{2}$, which dominates at small separations $h$, and (2) a `disconnected' hypersurface consisting of separate components for $A_{1}$ and $A_{2}$, which becomes minimal when $h$ exceeds a critical value. At this critical separation, $S_{A_{1} \cup A_{2}} = S_{A{1}} + S_{A{2}}$; indicating the vanishing of mutual information.

We choose the separation $h$ to be very small relative to the strip-width $\ell$ -- this ensures that connected extremal hypersurface always has the minimum area, and mutual information is positive throughout the evolution.

\begin{figure}[t]
	\centering
	\begin{subfigure}{0.48\textwidth}
		\includegraphics[width=\textwidth]{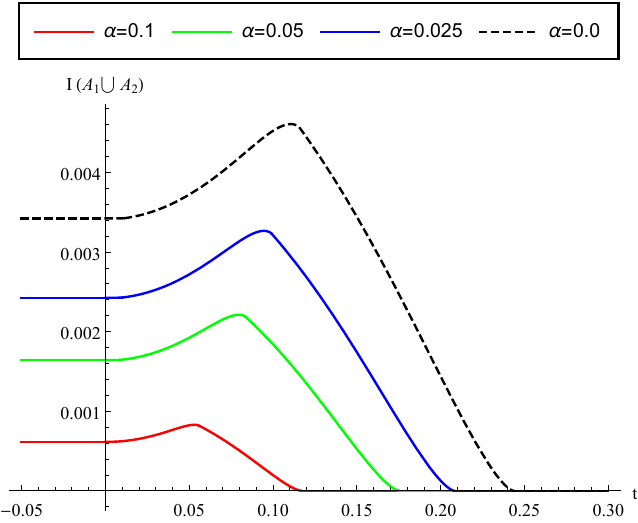} 
		\caption{} 
		\label{MI instant} 
	\end{subfigure}
	\begin{subfigure}{0.48\textwidth}
		\includegraphics[width=\textwidth]{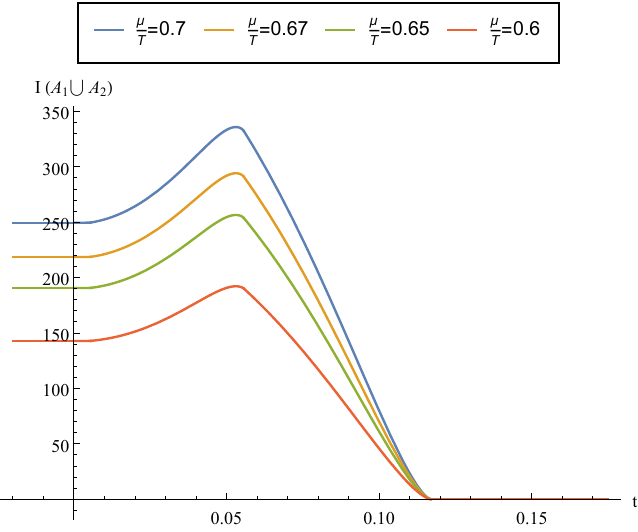} 
		\caption{} 
		\label{MI instant mute} 
	\end{subfigure}
	\caption{Evolution of mutual information $I\left(A_{1} \cup A_{2} \right)$ for (a) different Gauss-Bonnet coupling, and (b) different $\frac{\mu}{T}$ ratio.}
\end{figure}

Previous studies have demonstrated that the holographic mutual information exhibits a non-monotonic behaviour as a function of time \cite{Balasubramanian:2011at, Allais:2011ys, Callan:2012ip, Li:2013sia, Ziogas:2015aja}. Our analysis reveals a qualitatively similar evolution of HMI. Specifically, for fixed subregion sizes and separations, we investigate the effect of varying the Gauss-Bonnet coupling ($\alpha$) on HMI. We observe that for all types of quenches considered -- instantaneous, linear, and periodic -- the holographic mutual information is initially positive, starts growing after the quench, attains a maximum, and finally it decays to its saturation value determined by the black hole in the final state. Physically this is easy to explain: before the quench $I\left(A_{1} \cup A_{2} \right)$ is the same as in an empty AdS$_{5}$ background, immediately after the quench some energy is pumped into the system and the correlation between $A_{1}$ and $A_{2}$ starts to grow. As the system reaches equilibrium, the amount of correlation in the final state also gets determined by the equilibrium parameters and saturates. We find that increasing the Gauss-Bonnet coupling leads to a suppression of HMI, consistent with the behaviour observed in the holographic entanglement entropy. These observations are illustrated in figure \ref{MI instant} for an instantaneous quench. Let us clarify that mutual information between the subregions in the equilibrium state is not zero. In these plots, we actually subtracted the final equilibrium value of mutual information for each $\alpha$. 

We can also check how changing the chemical potential $\left(\mu\right)$ and the temperature $\left(T\right)$ affects the change in mutual information. In figure \ref{MI instant mute} we plot the time evolution of $I\left(A_{1} \cup A_{2} \right)$ for different $\frac{\mu}{T}$ ratio. As we have already seen in the case of HEE, these two parameters do not play any role in the time scales involved.

Mutual information for the other two kinds of quenches can be plotted easily. We do not include the figures as the conclusions are the same.

\section{Concluding remarks} \label{conclusions}

In this work, we investigated the evolution of holographic entanglement entropy in Einstein-Gauss-Bonnet gravity following a global quench. By considering different quench profiles, we analyzed the entanglement growth for both strip and spherical subsystems and identified universal features in the thermalization process. Our findings indicate that the Gauss-Bonnet coupling influences the saturation time and modifies the rate of entanglement propagation, yet the overall qualitative behaviour remains consistent with known holographic results.

We also explored the linear response regime, where the deviation from equilibrium can be captured by a time-dependent relative entropy. Our results support the robustness of the first law-like behaviour for slowly varying quenches and provide insights into the role of higher-derivative corrections in holographic thermalization. Further extensions, such as including effects of non-zero angular momentum or studying other kinds of non-equilibrium phenomenon in holographic higher-derivative gravity, would be interesting avenues for future research.

Although in this work we worked in a limited region of the allowed parameter space where both the Gauss-Bonnet coupling $\alpha$ and the strip-width $\ell$ were considered to be very small, we are of the opinion that even for larger $\alpha$ our conclusions about the thermalization of entanglement entropy for small subsystems would remain equally valid. For large subsystems, the results will change and different growth regimes are expected to arise.

\section*{Acknowledgments}
SP would like to thank Pabitra Tripathy and Nemai Chandra Sarkar for useful discussions and help with Mathematica. SP acknowledges financial support from Saha Institute of Nuclear Physics (SINP), Department of Atomic Energy (DAE), Govt. of India. 

\appendix

\section{Holographic entanglement entropy for Gauss-Bonnet gravity in other dimensions} \label{appendix:other_d}

In the main text, we studied the holographic entanglement entropy in detail for Einstein-Gauss-Bonnet gravity in $\left(4+1\right)$ spacetime dimensions. The analysis can be easily replicated for other dimensions, and in this appendix we will summarise the results in six and seven spacetime dimensions. We will only focus on the strip subregion, as the results are qualitatively same for both the strip and the ball.

\subsection*{Entanglement entropy in $\left(5+1\right)$ dimensions}

In the context of six dimensional charged Gauss-Bonnet black brane, using the setup described in section \ref{sec2} and \ref{subsec 3.2}, the area functional for the holographic entanglement entropy of a strip can be expressed as
\begin{equation}
	\mathcal{A}(t)=2 L^3_{\text{eff}}\ell^3_{\perp}\int_{0}^{z_{*}}dz \frac{\sqrt{x'^2-F(v,z)v'^2-2v'}}{z^4}\left(1+\frac{12\alpha}{x'^2-F(v,z)v'^2-2v'}\right)
\end{equation}
where prime denotes derivative w. r. t. z. The limits of integration are taken from the boundary (z=0) till a turning point $z=z_{*}$, where $x(z_{*})=0$.

To obtain the equations of motion, we make use of the same small subregion approximation $\lambda = \frac{z_{*}}{z_{h}} \ll 1$ as well as assume $\alpha \ll 1$. The Euler-Lagrange equation for the extremal hypersurface is
\begin{equation} \label{eq:EOM_6d}
	 \frac{x(z)}{z^4 \sqrt{1 + x(z)^2}}-\frac{9 \alpha x(z)}{z^4 \left(1 + x(z)^2\right)^{3/2}}=\text{constant}. 
\end{equation}
The constant on the right-hand side is determined by the condition \( x'(z) \to \infty \) as \( z \to 1 \). Using this, Equation \eqref{eq:EOM_6d} can be solved to derive a perturbative solution for the hypersurface \( x(z) \) in a static, pure anti-de Sitter spacetime at \( \mathcal{O}(\alpha) \).
\begin{equation}
	x(z) \simeq \frac{\ell}{2} \left(1-3\alpha \right) - \frac{1}{5} z^5 \, {}_2F_1\left(\frac{1}{2}, \frac{5}{8}, \frac{13}{8}, z^8\right)(1+9\alpha) + \mathcal{O}\left(\alpha^2 \right).
\end{equation}
The relationship between the turning point \( z_* \) and the strip width \( \ell \) can be readily determined using the appropriate formula
\begin{equation}
	z_{*}=\frac{\ell}{2}\frac{ \, \Gamma\left(\frac{1}{8}\right)}{\sqrt{\pi} \, \Gamma\left(\frac{5}{8}\right)} \left(1-12\alpha \right) + \mathcal{O} \left(\alpha^2 \right).
\end{equation}
We use this solution to calculate the leading-order change in holographic entanglement entropy by subtracting the zeroth-order terms. For simplicity, let us only consider the instantaneous quench. It is pretty straightforward to generalize the calculation for the other two types of quenches, and the conclusions remain qualitatively identical. For an instantaneous quench, we find
\begin{equation}
	\Delta S_{A}(t)=\frac{L_{\mathrm{eff}}^3 \ell_{\perp}^3 \lambda^5\varepsilon}{4 G_N} \int_0^1 dz \,z\sqrt{1-z^8}(1-9\alpha)\Theta\left(t-\frac{z}{\frac{1 - \sqrt{1 - 24\alpha}}{12\alpha}}\right) 
\end{equation}
\begin{equation}
	\simeq\frac{ \ell_{\perp}^3 \lambda^5\varepsilon}{4 G_N} \int_0^1 dz \,z\sqrt{1-z^8}(1-18\alpha)\Theta\left(t-(1-6\alpha)z\right) 
\end{equation}
We can evaluate this integral in three different regimes, as done previously, and the key result of the calculation is as follows
\begin{equation} \label{eq:Delta_S_6D}
	\Delta S_{A}(t) = \Delta S_{\text{eq}}\{\left[\Theta(t) - \Theta\left(t - t_{\text{sat}} \right) \right] \mathcal{G}\left(\frac{t}{t_{\text{sat}}} \right) + \Theta \left(t - t_{\text{sat}} \right)\},
\end{equation}
Where
\begin{align}
	t_{\text{sat}} &= (1-6\alpha)z_{*},\\
	\Delta S_{eq} &= \frac{\varepsilon\ell^3_{\perp}}{4G_{N}}\frac{z^2_{*}}{z^5_h}\frac{\sqrt{\pi} \, \Gamma\left(\frac{1}{4}\right)}{12 \, \Gamma\left(\frac{3}{4}\right)}(1-18\alpha),
\end{align}
and $\mathcal{G}(x)$ is given by
\begin{equation}
	\mathcal{G}(x) = \frac{6\Gamma\left(\frac{3}{4}\right)}{\sqrt{\pi}\Gamma\left(\frac{1}{4}\right)} x^2 \, {}_2F_1\left(-\frac{1}{2}, \frac{1}{4}; \frac{5}{4}; x^8\right), \quad x = \frac{t}{t_{\text{sat}}}.
\end{equation}
We can also calculate the instantaneous rate of change of the entanglement growth,
\begin{equation}
	\mathcal{R}(t)=\frac{\ell}{2 t_{\text{sat}}}\frac{d\mathcal{G}}{dx}=\frac{\ell}{t_{\text{sat}}}\frac{6\Gamma\left(\frac{3}{4}\right)}{\sqrt{\pi}\Gamma\left(\frac{1}{4}\right)}\frac{t}{t_{\text{sat}}}\sqrt{1-\left(\frac{t}{t_{\text{sat}}}\right)^8}
\end{equation}
and the time averaged entanglement velocity
\begin{equation}
	v^{\text{avg}}_{E}=\langle \mathcal{R}(t) \rangle=\frac{\ell}{2 t_{\text{sat}}}=\frac{\sqrt{\pi}\Gamma(\frac{5}{8})}{\Gamma(\frac{1}{8})}(1+18\alpha)+\mathcal{O}(\alpha^2)
\end{equation}
\begin{figure}[t]
	\centering
	\begin{subfigure}[b]{0.48\textwidth}
		\centering  
		\includegraphics[width=\textwidth]{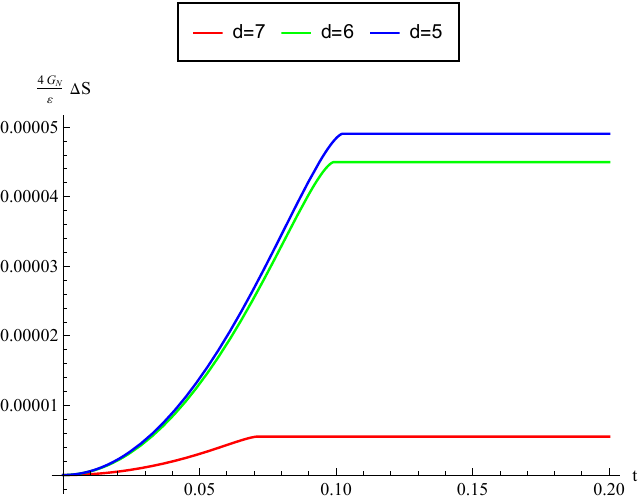}
		\caption{}
		\label{fig:HEE_strip_dims}
	\end{subfigure}
	\hfill
	\begin{subfigure}[b]{0.48\textwidth}
		\centering  
		\includegraphics[width=\textwidth]{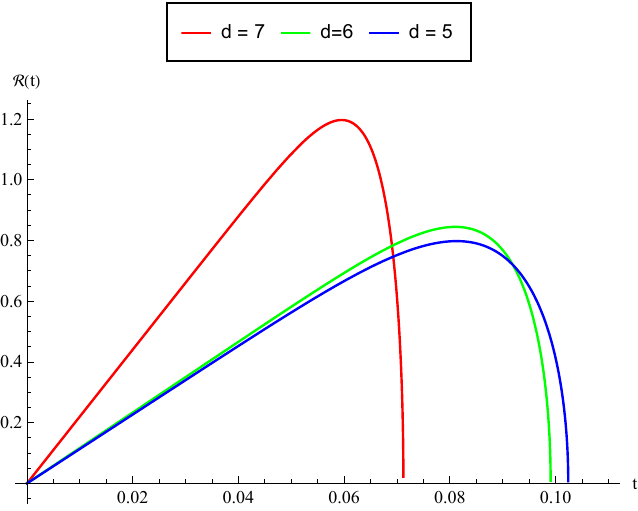}
		\caption{}
		\label{fig:Rate_strip_dims}
	\end{subfigure}
	\caption{Evolution of holographic entanglement entropy for a strip in  Gauss-Bonnet gravity with a fixed Gauss-Bonnet coupling $\alpha=0.02$. Here we set $\frac{\ell}{z_{h}}=0.1$ for small subregion, and   ${\ell^2_{\perp}}=1$ for simplicity. In (a) we show the growth of HEE after an instantaneous global quench for different spacetime dimensions. In (b) we plot the instantaneous growth-rate $\mathcal{R}(t)$ of entanglement entropy.}
	\label{fig:Instantaneous_strip_dims}
\end{figure}

\subsection*{Entanglement entropy in $\left(6+1\right)$ dimensions}

The equation of motion in $\left(6+1\right)$ spacetime dimensions is given by
\begin{equation}\label{eq:EOM_7D}
	\frac{x(z)}{z^5 \sqrt{1 + x(z)^2}}-\frac{18 \alpha x(z)}{z^5 \left(1 + x(z)^2\right)^{3/2}} = \text{constant}, 
\end{equation}
with the following solution at $\mathcal{O}\left(\alpha\right)$
\begin{equation}
	x(z) = \left(1-6\alpha \right) \frac{\ell}{2}-\frac{1}{6} z^6 \, {}_2F_1\left(\frac{1}{2}, \frac{3}{5}, \frac{8}{5}, z^{10}\right) \left(1+18\alpha \right) + \mathcal{O} \left(\alpha^2 \right).
\end{equation}
The turning point in this case gets related to the strip-width $\ell$ by
\begin{equation}
	z_{*}=\frac{\ell}{2}\frac{ \, \Gamma\left(\frac{1}{10}\right)}{\sqrt{\pi} \, \Gamma\left(\frac{3}{5}\right)}(1-24\alpha)+\mathcal{O}(\alpha^2)
\end{equation}
Using these information, we calculate the first order change in holographic entanglement entropy
\begin{equation}
	\Delta S_{A}(t) = \frac{ \ell_{\perp}^4 \lambda^6\varepsilon}{4 G_N} \int_0^1 dz \,z\sqrt{1-z^{10}}(1-42\alpha)\, \Theta\left(t-\left(1-12\alpha\right)z\right). 
\end{equation}
Solving the integral as before, we arrive at
\begin{equation} \label{eq:Delta_S_7D}
	\Delta S_{A}(t) = \Delta S_{\text{eq}}\{\left[\Theta(t) - \Theta\left(t - t_{\text{sat}} \right) \right] \mathcal{H}\left(\frac{t}{t_{\text{sat}}} \right) + \Theta \left(t - t_{\text{sat}} \right)\},
\end{equation}
where
\begin{align}
	t_{\text{sat}} &=(1-12\alpha)z_{*},\\
	\Delta S_{eq} &= \frac{\varepsilon\ell^4_{\perp}}{4G_{N}}\frac{z^2_{*}}{z^6_h}\frac{\sqrt{\pi} \, \Gamma\left(\frac{1}{5}\right)}{14 \, \Gamma\left(\frac{7}{10}\right)}(1-42\alpha),
\end{align}
and $\mathcal{H}(x)$ is given by
\begin{equation}
	\mathcal{H}(x) = \frac{7\Gamma\left(\frac{7}{10}\right)}{\sqrt{\pi}\Gamma\left(\frac{1}{5}\right)} x^2 \, {}_2F_1\left(-\frac{1}{2}, \frac{1}{5}; \frac{6}{5}; x^{10}\right), \quad x = \frac{t}{t_{\text{sat}}}.
\end{equation}
We can also calculate the instantaneous rate of change of the entanglement growth,
\begin{equation}
	\mathcal{R}(t)=\frac{\ell}{2 t_{\text{sat}}}\frac{d\mathcal{H}}{dx}=\frac{\ell}{t_{\text{sat}}}\frac{7\Gamma\left(\frac{7}{10}\right)}{\sqrt{\pi}\Gamma\left(\frac{1}{5}\right)}\frac{t}{t_{\text{sat}}}\sqrt{1-\left(\frac{t}{t_{\text{sat}}}\right)^{10}}
\end{equation}
and the time averaged entanglement velocity
\begin{equation}
	v^{\text{avg}}_{E}=\langle \mathcal{R}(t) \rangle=\frac{\ell}{2 t_{\text{sat}}}=\frac{\sqrt{\pi}\Gamma(\frac{3}{5})}{\Gamma(\frac{1}{10})}(1+36\alpha)+\mathcal{O}(\alpha^2)
\end{equation}

The time-evolution of entanglement entropy $\Delta S_{A}(t)$ and its instantaneous rate of growth $\mathcal{R}(t)$ in different dimensions is depicted in figure \ref{fig:Instantaneous_strip_dims}.

\bibliographystyle{JHEP}
\bibliography{references}

\end{document}